\documentclass[aps, pra, reprint, lengthcheck, footinbib, showpacs, superscriptaddress, citeautoscript]{revtex4-2}

\usepackage[T1]{fontenc}

\usepackage{amsmath}

\usepackage{braket}
\usepackage{graphicx}

\usepackage[colorlinks, urlcolor=blue, citecolor=blue, linkcolor=blue, pdfstartview=FitH]{hyperref}
\usepackage[all]{hypcap}

\usepackage{dcolumn}
\usepackage{bm}
\usepackage{epstopdf}

\usepackage{color}
\usepackage{xcolor}
\usepackage[normalem]{ulem}
\definecolor{color}{rgb}{0.11,0.45,0.02}

\begin{document}
	
\def\affiE2a{Experimentelle Physik 2, Technische Universit\"at Dortmund, 44221 Dortmund, Germany}
\def\affiIOFFE{Ioffe Institute, Russian Academy of Sciences, 194021 St.~Petersburg, Russia}
\def\affiParis{Laboratoire de Physique et d'\'{e}tude des
	Mat\'{e}riaux, ESPCI, CNRS, 75231 Paris, France}

\title{Optical alignment and orientation of excitons in ensemble of core/shell CdSe/CdS colloidal nanoplatelets}

\author{O.~O.~Smirnova}
\affiliation{\affiIOFFE}
\author{I.~V.~Kalitukha}
\affiliation{\affiIOFFE}
\author{A.~V.~Rodina}
\affiliation{\affiIOFFE}
\author{ G.~S.~Dimitriev}
\affiliation{\affiIOFFE}
\author{V.~F.~Sapega}
\affiliation{\affiIOFFE}
\author{O.~S.~Ken}
\affiliation{\affiIOFFE}
\author{V.~L.~Korenev}
\affiliation{\affiIOFFE}
\author{N.~V.~Kozyrev}
\affiliation{\affiIOFFE}
\author{S.~V.~Nekrasov}
\affiliation{\affiIOFFE}
\author{Yu.~G.~Kusrayev}
\affiliation{\affiIOFFE} 
\author{D.~R.~Yakovlev}
\affiliation{\affiE2a}
\affiliation{\affiIOFFE}
\author{B.~Dubertret}
\affiliation{	\affiParis}
\author{M.~Bayer}
\affiliation{\affiE2a}

\date{\today}

\begin{abstract}
\textbf{Abstract:} 
We report on the experimental and theoretical studies of optical alignment and optical orientation effects in an ensemble of core/shell CdSe/CdS colloidal nanoplatelets. The dependences of three Stokes parameters on the magnetic field applied in the Faraday geometry are measured under continuous wave resonant excitation of the exciton photoluminescence. Theoretical model is developed to take into account both bright and dark exciton states in the case of strong electron and hole exchange interaction and random in-plane orientation of the nanoplatelets in ensemble. The data analysis allows us to estimate the time and energy parameters of the bright and dark excitons. The optical alignment effect enables identification of the exciton and trion contributions to the photoluminescence spectrum even in the absence of a clear spectral line resolution.
\end{abstract}

\maketitle


Colloidal semiconductor nanocrystals are of interest for various fields of chemistry, physics, biology, and medicine and are successfully used in various optoelectronic devices. Being synthesized from many different semiconductor materials, they can have different geometries, resulting in zero-dimensional quantum dots, one-dimensional nanorods, or two-dimensional nanoplatelets (NPLs). CdSe NPLs demonstrated small inhomogeneous broadening \cite{Ithurria2008}, as confirmed later also for NPLs of different composition \cite{Nasilowski2016,Berends2017,Ithurria2012}. Semiconductor NPLs can be considered as model systems to study physics in two dimensions, similar to quantum well heterostructures and layered materials, like graphene or transition-metal dichalcogenides.

However, the spin physics of colloidal nanocrystals is still in its infancy compared to the rather mature field of spintronics based on epitaxially grown semiconductor quantum wells and quantum dots. The properties of colloidal and epitaxial nanostructures can differ considerably due to much stronger confinement of charge carriers in colloidal structures, leading to different properties such as the strongly enhanced Coulomb interaction (enhanced exciton binding energy and fine structure energy splitting), the possibility of photocharging and surface magnetism, the reduction of the phonon bath influence, etc. Experimental techniques widely used in spin physics of heterostructures can be, however, readily applied to
colloidal quantum dots and NPLs. Among them are polarized photoluminescence (PL) spectroscopy, including optical orientation and optical alignment methods.

Methods of optical orientation and optical alignment is used to study the population and coherence of different states in solids. Optical orientation is the excitation of states with a certain projection of angular momenta by circularly polarized light. In optics, optical orientation manifests itself in the polarization of the photoluminescence (PL). Optical alignment consists in the excitation of states with a certain direction of dipole moment by linearly polarized light \cite{Pikus_book1982}. In semiconductors, optical orientation and alignment can be observed for both electron-hole pairs and excitons.

Optical orientation of electron-hole pairs consists in creating of preferential populations of electron and hole sublevels with different spin projections~\cite{OO_book}. Their recombination results in a circularly polarized PL whether or not this pair was generated in the same act of light absorption. On the contrary, optical alignment involves the preservation of correlation (coherence) between the spins of the electron and hole in the pair. Such correlation is possible only for the recombination of an electron and a hole generated in one act of absorption of linearly polarized light \cite{Bir1976}.
Usually, however, electron and hole belonging to different pairs recombine, correlation is not preserved and the alignment effect is not observed for electron-hole pairs.

Under resonant excitation of the exciton, a bound electron-hole pair is generated, so both optical orientation and alignment are preserved. In the case of excitons in GaAs-type nanostructures \cite{IvchenkoPikus_book}, the optical orientation is given by the difference of populations of bright exciton states with a projection of the total angular momentum $+1$ and $-1$ due to the absorption of circularly polarized light. The optical alignment consists in creating a coherent superposition of a pair of optically active states $\pm1$ by linearly polarized light. Both the difference of populations and coherence evolve in time under the influence of magnetic interactions of different nature (Zeeman, exchange, etc.). If optical orientation and alignment of excitons preserve during their lifetime, the PL is partially polarized. In this case three Stokes parameters unambiguously determine the polarization state of the ensemble of bright excitons. Optical orientation and alignment are not independent. The conversion of exciton coherence to population and vice versa is observed \cite{Dzhioev1997PRB}.

The study of optical orientation and alignment provides comprehensive information on the fine structure of exciton spin levels. In a zero magnetic field, the ground state of the exciton in NPLs is quadruple degenerated. The exchange interaction between an electron and a hole splits this state into an optically active doublet (bright excitons) and two energetically close optically forbidden singlets (dark excitons). Exciton localization in the NPL lowers the symmetry of the system, and optically active doublet splits into two sublevels \cite{Goupalov1998_2,Dzhioev1997}, which are linearly polarized  in two orthogonal directions, whose orientation is given by the symmetry of the localizing potential. Knowledge of the fine structure of the exciton (the directions of the main axes and the values of splittings) allows one to infer the symmetry of the nanostructure and the selection rules, which is important for understanding the radiation efficiency and pattern of NPLs. Unfortunately, practically one often deal with an ensemble of NPLs various both in shape and size. In this case, characteristic splittings in the exciton fine structure are indistinguishable in the PL spectrum due to a large energy broadening of optical transitions. Nevertheless, the exciton fine structure is evident in the polarization of exciton PL even in the absence of spectral resolution \cite{IvchenkoPikus_book,Dzhioev1997PRB,Dzhioev1997,Dzhioev1998}.

Nanostructures are often doped or photocharged with electrons or holes. A significant contribution to the optical properties of such nanostructures is made by three-particle complexes -- trions, consisting of two electrons (or holes) in the singlet state and one unpaired hole (or electron). Trion PL overlaps the exciton PL and complicates the interpretation of optical orientation experiments \cite{Dzhioev1998_2}. Optical alignment becomes essential in this case. There is no correlation between electron and hole spins in the trion, and therefore there is no optical alignment effect. The presence of exciton optical alignment allows one to distinguish between the exciton contribution to the linear polarization even in the absence of a clear spectral resolution between excitons and trions.

Optical orientation and alignment spectroscopy have proven to be powerful tools for studying the spin structure in semiconductors. These methods contributed previously a lot to the characterization of electronic excitations in semiconductors: the times of dynamic processes \cite{EkimovSafarov1970, Ekimov1971_2, Gamarts1977, Dymnikov1981}, effective $g$-factors of carriers \cite{Dzhioev1973} and parameters of exciton fine structure \cite{Dzhioev1997PRB}. Optical orientation, optical alignment, and polarization conversion effects were observed in epitaxial nanostructures \cite{Kusrayev2008,Koudinov2008}. Among the family of colloidal semiconductor nanostructures (chemically synthesized in solution or dispersed in a dielectric matrix semiconductor nanocrystals  synthesized), there is a  report on the observation and theoretical description of the optical orientation and alignment in inorganic perovskite nanocrystals \cite{Nestoklon2018}.

In conventional II-VI semiconductor colloidal nanocrystals the optical orientation and alignment have not been studied so far. We observe optical alignment and optical orientation of bright and dark exciton in core/shell CdSe/CdS NPLs. We develop a theory of polarized PL in order to describe the effects taking into account the peculiarities associated with colloidal NPLs: the large electron-hole exchange interaction resulting in large bright and dark exciton splitting and the random in-plane orientation of the NPLs in the ensemble. The analysis of the experimental data allows us to estimate the anisotropic energy splitting in bright and dark exciton states in zero magnetic field, the $g$-factors of bright and dark excitons, and the bright and dark exciton pseudospin lifetimes.

\section{\label{sec:Experiment}Experiment}

\begin{figure}[h!]
	\centering
	\includegraphics[width=8.6 cm]{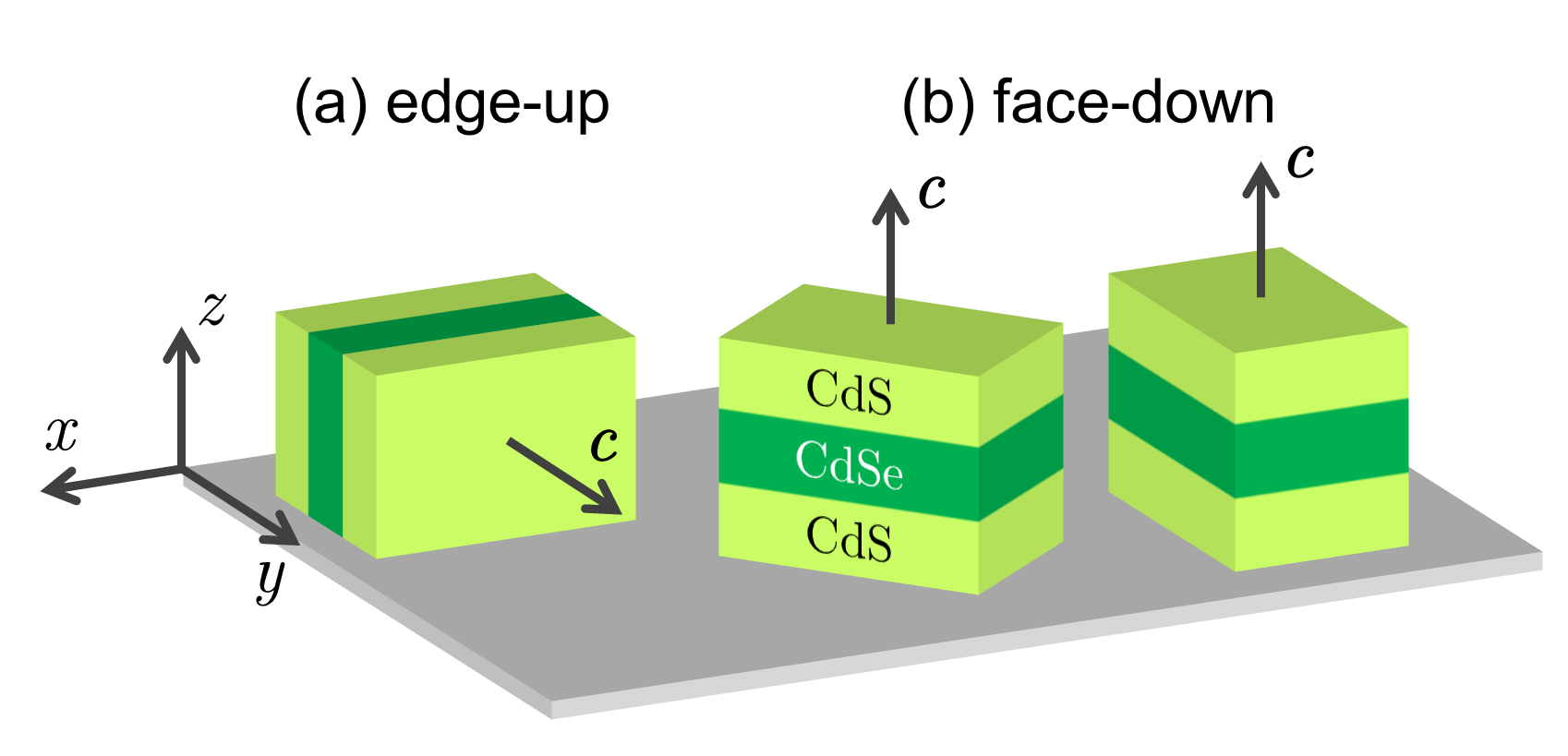}
	\caption{\label{fig:npl_geom} Preferred orientations of the NPLs on the substrate: (a) vertically oriented NPLs (edge-up) with ${\bm c}$ lying in the substrate plane and (b) horizontally oriented NPLs (face-down) with anisotropic NPL axis ${\bm c}$ perpendicular to the substrate.} 
\end{figure}

NPLs with 4 monolayers (MLs) CdSe core and CdS shell (schematically represented in Fig.~\ref{fig:npl_geom}) of different thicknesses were synthesized in the group of Benoit Dubertret in Paris, see Methods. The samples were investigated previously by means of a pump-orientation-probe technique, detecting the electron spin coherence at room temperature \cite{Feng2020}. In particular, the electron $g$-factor and its dependence on the shell thickness 
were reported.

In the present paper we conducted the low-temperature and  temperature-dependent studies of the PL properties of the NPLs with different shell thickness. We study the effects of the optical alignment and optical orientation of excitons, which were not observed
The effects of optical alignment and optical orientation are qualitatively the same for samples with a different shell thickness. We focus here on the NPLs with a medium CdS shell thickness of $3.1$~nm at each side of the CdSe core.
The PL spectrum of this sample at a temperature of $T=1.5$~K measured under either resonant $E_{\rm exc} =  1.960$~eV (Figure \ref{cw}(a) in Supplementary Information (SI)) or non-resonant $E_{\rm exc} =  2.380$~eV (Figure S2(a)) excitation  consists of one broad band. This is typical for CdSe/CdS  core/shell NPLs \cite{Shornikova2018nl,Shornikova2020acs,Shornikova2020nn}, in contrast to the bare core CdSe NPLs which PL spectrum has separated  exciton and trion lines \cite{Shornikova2020nl}. 


\begin{figure*}
	\centering
	\includegraphics[width=0.8\textwidth]{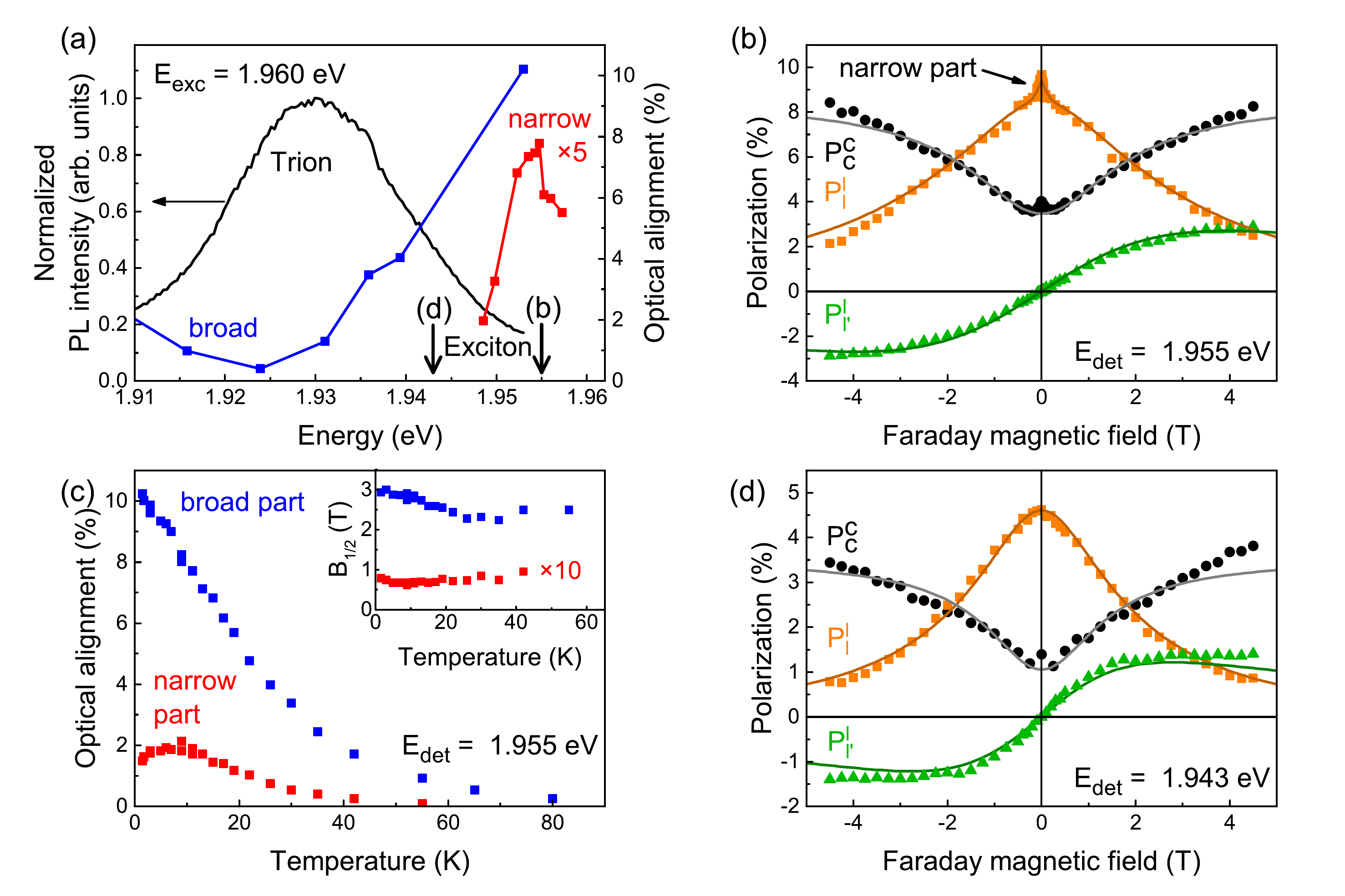}
	
	\caption{\label{cw} Polarized PL spectroscopy of CdSe/CdS NPLs under resonant excitation with the energy $E_{\rm exc} = 1.960$~eV. (a) PL spectrum (black), amplitudes of the broad (blue squares) and narrow (red squares, multiplied by 5) parts of the optical alignment contour. Arrows are pointing at the detection energies for panels (b) and (d). (b) Experimental data (symbols) and theoretical calculations (solid lines of corresponding colors) of the optical alignment (orange squares), the rotation of the linear polarization plane (green triangles) and the optical orientation (black circles) measured in the Faraday geometry at $E_{\rm det} = 1.955$~eV. (c) Temperature dependence of the amplitude of the broad (blue squares) and narrow (red squares) parts of the optical alignment contour. Inset: temperature dependence of their HWHM (denoted as $B_{1/2}$). $B_{1/2}$ of narrow part is multiplied by 10. (d) Experimental data (symbols) and model calculations (solid lines) of the optical alignment (orange squares), the rotation of the linear polarization plane (green triangles) and the optical orientation (black circles) measured in the Faraday geometry at $E_{\rm det} = 1.943$~eV.
	}
\end{figure*}

Polarization of light is characterized by three Stokes parameters. Two of them characterize linear polarization degree in $x$, $y$ axes and in $x'$, $y'$ axes rotated by 45$^{\circ}$ degrees around $z$ axis ($P_l$ and $P_{l'}$, respectively). The third one characterizes the circular polarization degree ($P_c$). For the light propagating along  $z$ direction polarizations are defined as 
\begin{equation}\label{key}
	P_{l} = \frac{I_x-I_y}{I_x+I_y},\ \ \  P_{l'} = \frac{I_{x'}-I_{y'}}{I_{x'}+I_{y'}}, \ \ \  P_c = \frac{I_+-I_-}{I_++I_-}. 
\end{equation}
Here $I_{x(y)}$ is the intensity of horizontally (vertically) polarized component, $I_{x'(y')}$ is the intensity of 45$^{\circ}$ ($-45^{\circ}$) polarized component, and $I_{+(-)}$ is the intensity of $\sigma^+$ ($\sigma^-$) polarized component of light. 


In general, one can perform nine measurements, taking into account the combination of three options for the polarization of the incident and detected light. We denote each measurement as $P_{\alpha}^{\beta}$, which combines $\alpha$-polarized excitation $P_{\alpha}^0$ and $\beta$-polarized detection $P_{\beta}$ with $\alpha, \beta = c, l,l'$. For these samples due to random in-plane orientation of the NPLs on the substrate the optical alignment is independent of specific linear polarization of the exciting light ($P_{l}^l \sim P_{l'}^{l'}$), which is also true for the effect of rotation of the linear polarization plane ($P_{l'}^l \sim P_{l}^{l'}$). For the same reason, all the effects associated with the conversion of initial linear polarization to circular polarization and {\it vice versa} are absent, as evidenced by the zero experimental dependence on the magnetic field $P_c^l(B) = P_l^c(B) = P_c^{l'}(B) = P_{l'}^c(B) = 0$. Therefore, to describe polarization-dependent effects it is sufficient to investigate three nontrivial results: the effect of the optical orientation, measured as $P_c^c$, the effect of the optical alignment, measured as $P_l^l$, and the effect of the rotation of the linear polarization plane, measured as $P_{l'}^l$. While $P_l^l$ and $P_c^c$ are present in the absence of magnetic field, $P_{l'}^{l}$ manifests itself only in the nonzero magnetic field in the Faraday geometry. We study these three Stokes parameters in the magnetic fields up to $4.5$~T applied in the Faraday geometry with ${\bm B} \parallel {\bm k}$, where $ {\bm k}$ is wave vector of the incident light directed perpendicular to the substrate (and opposite to the wave vector of the emitting light). However, we keep the definition of the circular polarization sign $P_c^0$ and $P_c$ with the respect to the same direction of $z$.


\subsection{Optical alignment  and optical orientation in continuous wave experiment}
\label{section:cw}

The optical orientation, $P_c^c$, and optical alignment, $P_l^l$, are observed only under the resonant excitation of excitons. While the $P_c^c$ is observed all over the PL spectrum, the  $P_l^l$  depends crucially on detection energy. Figure \ref{cw}(a) shows the spectrum of the optical alignment (blue) under linearly polarized continuous wave (cw) excitation with the laser energy $E_{\rm exc}=1.960$~eV in comparison with the PL spectrum (black). The optical alignment effect has minimum at the PL maximum and increases toward the high-energy part of the spectrum up to $P_l^l=10\%$. This spectral dependence suggests that the PL at the high-energy part of the spectrum is contributed by the excitons. The PL around its maximum is mostly contributed by negative trions, composed of two electrons and one hole, because the optical alignment is not expected for singlet trions or any fermion quasi-particles. We verify the exciton origin of the PL  at the high-energy part of the spectrum by the PL time-resolved measurements (the results are discussed in the subsection \ref{section:TRPL}).

For ensemble of NPLs the two preferred orientations of the NPL anisotropic axis ${\bm c}$ with respect to the substrate are expected: vertically oriented NPLs (edge-up, Figure \ref{fig:npl_geom}(a)) with ${\bm c}$ lying in the substrate plane and horizontally oriented NPLs (face-down, Figure \ref{fig:npl_geom}(b)) with ${\bm c}$ perpendicular to the substrate \cite{Shornikova2020nn,Shornikova2018nl}. The edge-up NPLs emit linearly polarized light both under linear and circular polarized excitation ($\bm k \parallel z$). However, the contribution of the PL from these edge-up NPLs to $P_l^l$ is constant under the application of the magnetic field in the Faraday geometry allowing for subtraction it from the experimental data and refer to the field-dependent $P_l^l$ as to the effect of the optical alignment of excitons in the face-down NPLs.

Figure \ref{cw}(b) shows the dependence of the optical alignment $P_l^l$ (orange squares) on the magnetic field in the Faraday geometry at  $E_{\rm det} = 1.955$~eV. The optical alignment contour consists of two parts: the broad one has HWHM (half width at half maximum, denoted as $B_{1/2}$) about $3$~T and the narrow one -- less than $0.1$~T. The spectral dependences of the amplitudes of both parts of the $P_l^l$ contour are shown in Figure \ref{cw}(a) with blue (broad part) and red (narrow part) symbols. The narrow part of the optical alignment contour is resonant with respect to the laser excitation energy: the effect amplitude reaches the maximum when the PL is detected about 5~meV below the laser energy regardless of the particular laser energy in a reasonable range and rapidly vanishes with varying of detection energy.

Figure \ref{cw}(c) shows the temperature dependence of broad (blue squares) and narrow (red squares) parts of the optical alignment. The dependences of the amplitudes are presented on the main panel. Broad part is $10\%$ at $T=1.5$~K, decrease and disappear above $80$~K. Narrow part of the optical alignment contour has a dip in the region under $5$~K, $2\%$ amplitude around $T=5$~K, and then decrease similar to the broad part and disappear above $60$~K. Dependences of HWHMs are in the inset in Figure \ref{cw}(c). HWHM for both contributions have weak temperature dependence.

Figure \ref{cw}(b) also shows the effect of the rotation of the linear polarization plane $P_{l'}^l(B)$ (green triangles), which results in nonlinear antisymmetric dependence on the magnetic field applied in the Faraday geometry. The value of linear polarization degree in the magnetic field of $4$~T is $3\%$. In addition, Figure \ref{cw}(b) includes the recovery of the optical orientation $P_c^c(B)$ (black circles), which manifests itself in increasing in the magnetic field. The amplitude of the polarization recovery curve is $6\%$. Both effects have a characteristic field of $3$~T corresponding to the broad part of the optical alignment contour.

The magnetic field dependences of all three Stokes parameters -- $P_l^l$, $P_{l'}^l$ and $P_c^c$ -- measured at the detection energy $E_{\rm det} =  1.943$~eV are presented in Figure \ref{cw}(d). This energy is $17$~meV lower than the excitation energy, thus the narrow part of the optical alignment contour is absent here. The amplitudes of the optical alignment and the polarization recovery curve decreased to $4.5\%$ and $3\%$, respectively.
The characteristic value of the rotation of the linear polarization plane decrease to $1.5\%$ in $B = 4$~T. The characteristic  magnetic field of all three dependences remains the same of $3$~T in comparison with Figure \ref{cw}(b).

\subsection{Raman scattering}
\label{section:SRFS}

\begin{figure*}
	\centering
	\includegraphics[width=0.8\textwidth]{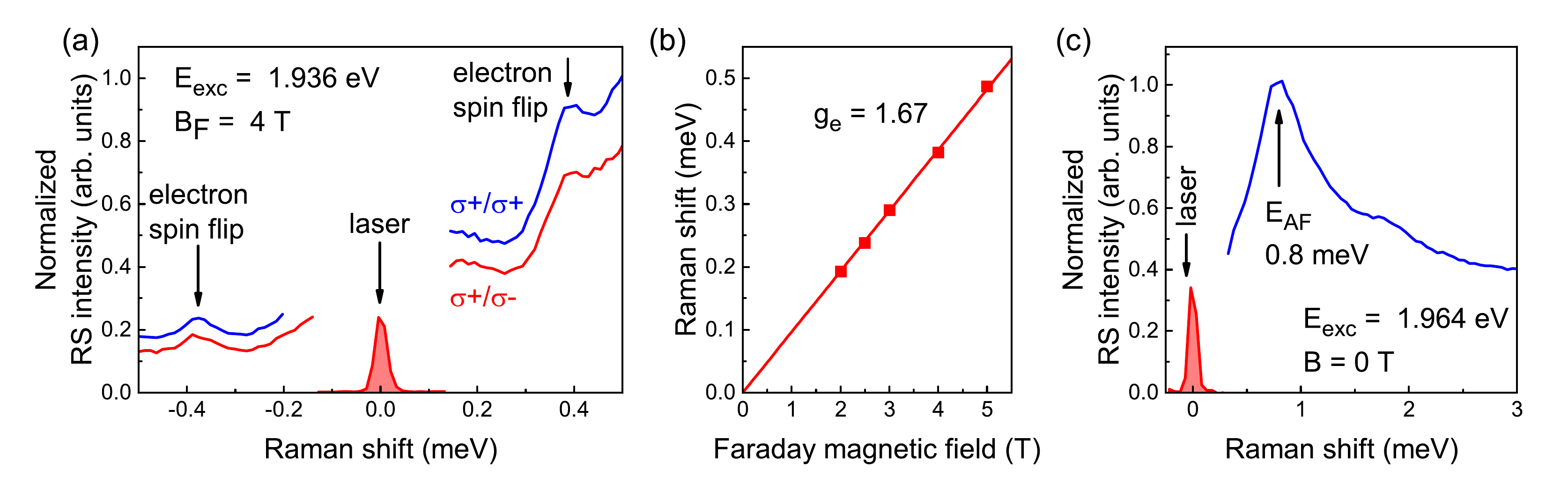}
	
	\caption{\label{SFRS} Raman scattering spectroscopy of CdSe/CdS NPLs. (a) SFRS spectra received under $E_{\rm exc}=1.936$~eV in the Faraday magnetic field $B_{\rm F}=4$~T at $T = 1.5$~K. Both spectra are measured under $\sigma^+$ polarized excitation. Blue (red) spectrum is $\sigma^+$($\sigma^-$) polarized component of measured signal, correspondingly. (b) Experimental dependence of Raman shift of electron spin flip line on magnetic field applied in the Faraday geometry (symbols) and its linear fit (line). (c) Raman scattering spectrum under excitation energy $E_{\rm exc} = 1.964$~eV at zero magnetic field at $T = 1.5$~K. Laser is shown by red filled curve.
	}
\end{figure*}

Determination of the electron $g$-factor $g_e$ is done by means of spin-flip Raman scattering (SFRS) spectroscopy. SFRS spectra in  co- and cross- circular polarizations received under resonant excitation at $E_{\rm exc}=1.936$~eV in the Faraday magnetic field $B_{\rm F}=4$~T at $T = 1.5$~K are shown in Figure \ref{SFRS}(a).

Both Stokes (with positive Raman shifts) and anti-Stokes (with negative Raman shifts) regions of spectra exhibit line at the energy $0.38$~meV. This line is attributed to the spin flip of the electron. Figure \ref{SFRS}(b) shows the magnetic field dependence of Raman shift of electron spin flip line. Its approximation with Zeeman energy $\mu g_e B_{\rm F}$ gives the $g$-factor of electron $g_e=1.67$. This value is close to the value $1.59$ determined for this sample at room temperature in \cite{Feng2020}. 

Raman scattering spectroscopy in zero magnetic field allows one to measure the energy splitting between the bright (optically allowed) and the dark (optically forbidden) exciton states. Figure \ref{SFRS}(c) shows Raman spectrum at $E_{\rm exc}=1.964$~eV at $T = 1.5$~K. The spectrum shows a line at energy $0.8$~meV. This line corresponds to the emission of the dark exciton after energy relaxation from the initially excited bright state. Thus, the energy splitting between the bright and the dark exciton states $\Delta E_{AF}=0.8$~meV.

\subsection{Time-resolved PL}
\label{section:TRPL}

The analysis of the temperature-dependent and magnetic-field dependent  PL decays allows us to prove that the high energy part of the spectrum (Figure \ref{cw}(a)) is contributed mostly by the exciton recombination. At the same time, trion determines dynamics at the PL maximum.

The PL spectra under pulsed excitation (both resonant and non-resonant) at $T = 1.5$~K consist of a broad line with maximum at $1.930$~eV, similar to the cw PL spectrum. It is shown in Supplementary Information (Figure S2(a)).

\begin{figure*}
	\includegraphics[width=0.8\linewidth]{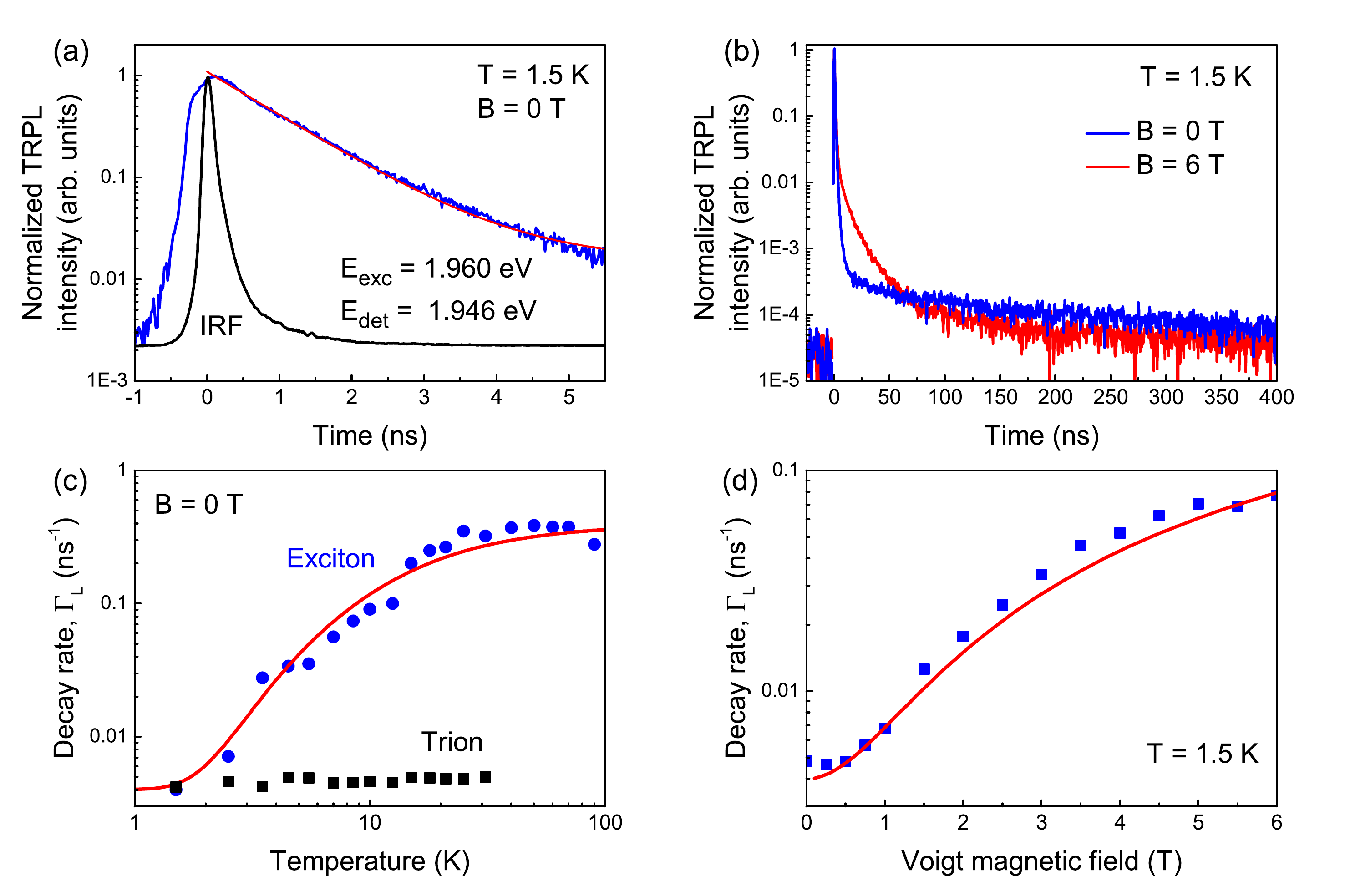}
	\caption{\label{TRPL} {Time-resolved PL spectroscopy of CdSe/CdS NPLs under resonant excitation with the energy $E_{\rm exc} = 1.960$~eV. (a) PL dynamics at short times (blue) with monoexponential fit (red). Black curve represents instrumental response function (IRF). (b) PL dynamics in different magnetic fields applied in the Voigt geometry. (c) Experimental temperature dependence of trion decay rate (black squares) and exciton long decay rate $\Gamma_{\rm L}$ (blue circles) with calculated one by Eq. \eqref{Gamma_As} (red line). (d) Experimental dependence of the asymptotic rate $\Gamma_{\rm L}$ on the magnetic field (blue squares) applied in the Voigt geometry and the theoretical curve (red solid line).}}
\end{figure*}

The dynamics of the PL obtained under resonant pulsed excitation with $E_{\rm exc} = 1.960$~eV and detected at the high-energy part of the PL spectrum ($E_{\rm det} = 1.946$~eV) is shown in Figure~\ref{TRPL}(a,b). These excitation-detection conditions correspond to the exciton PL. Exciton PL decay has short and long contributions. The short decay (within nanoseconds) is shown in Figure~\ref{TRPL}(a). It corresponds to the bright (optically allowed, A) exciton dynamics \cite{Shornikova2018ns} with the lifetime $\tau_A = 0.9$~ns. The long decay in zero magnetic field is shown in Figure~\ref{TRPL}(b) (blue curve). This component, characterized by a decay rate $\Gamma_{\rm L}$, is observed at $T=1.5$~K due to the population of the bright exciton state and its admixture to the dark (optically forbidden, F) exciton state.

PL decay is measured for the temperature range up to $90$~K (typical data are shown in Figure~S3(a)). The excitation-detection conditions corresponding to the resonant excitation of the exciton PL are the same for all the temperatures since the PL spectrum practically does not shift or change the shape with temperature (Figure~S2(a)). Figure~\ref{TRPL}(c) shows the temperature dependence of the decay rate $\Gamma_{\rm L}$. Such a behavior is typical for the exciton recombination \cite{Shornikova2018ns}. It is caused by the temperature-induced redistribution of the bright (characterised by the recombination rate $\Gamma_A$) and the dark (characterised by $\Gamma_F$) exciton states populations (see SI section S1). The analysis of the $\Gamma_{\rm L}(T)$ allows one to determine the exciton parameters \cite{Shornikova2018ns} including the bright and dark exciton recombination rates, as well as  the relaxation rates and the energy splitting between the bright and the dark exciton states $\Delta E_{AF}$ caused by the electron-hole exchange interaction (see SI, S1). Already mentioned bright exciton low-temperature lifetime $\tau_A = 0.9$~ns can be also expressed via $\tau_A = (\Gamma_A + \gamma_0)^{-1}$, where $\gamma_0$ gives the relaxation rate from bright to dark exciton \cite{Shornikova2018ns}. In conjunction with this, the  fitting of the $\Gamma_{\rm L}(T)$ dependence with Eq. \eqref{Gamma_As} allows us to determine $\gamma_0 = 0.31$~ns$^{-1}$, $\Gamma_A=0.8$~ns$^{-1}$, $\Gamma_F=0.004$~ns$^{-1}$ and $\Delta E_{AF} = 0.8$~meV.

The PL decay depends also on the magnetic field applied in the Voigt geometry (Figure~\ref{TRPL}(b)). The dependence of $\Gamma_{\rm L}(B)$, shown in Figure~\ref{TRPL}(d), is caused by the field-induced mixing of the bright and the dark exciton states. Early determined exciton parameters allow us also to describe $\Gamma_{\rm L}(B)$ via Eq. \eqref{Gamma_As}.

Temperature series of the PL decay are also recorded under non-resonant excitation energy $E_{\rm exc} = 2.380$~eV (Figure~S3(c)) with $E_{\rm det} = 1.952$~eV. $\Gamma_{\rm L}(T)$ dependence coincides with the case of the resonant excitation (Figure~S3(d)). Thus, independently of the excitation energy and in some range of the detection energies, the PL around $1.946$~eV (where the effect of the optical alignment is observed, see Figure \ref{cw}(a)), is contributed mostly by the exciton recombination. In contrast, the trion PL dynamics at $1.930$~eV (Figure~S3(b)) does not depend on the temperature (Figure~\ref{TRPL}(c)). This is typical for trions which do nor have dark state \cite{Shornikova2018nl,Shornikova2020nl}.

The electron-hole exchange interaction, as well as the direct Coulomb interaction in NPLs, is enhanced by the strong spatial confinement along ${\bm c}$ axis and by the dielectric contrast between the NPL and its surrounding ligands \cite{Shornikova2018ns,Shornikova2021}. As a result, in bare core $4$~ML CdSe NPLs $\Delta E_{AF}$ and the excition binding energy reach $4.5$~meV and $270$~meV, correspondingly \cite{Shornikova2021}.  The presence of the CdS shell decreases the electron-hole overlap because of the leakage of the significant part of the electron wave function into the shell and also decreases influence of the dielectric contrast. This results into decrease of  $\Delta E_{AF}$ to $0.8$~meV in the studied NPLs and the respective decrease of the exciton binding energy. However, the latter remains large excluding the possibility to excite the unbound electron-hole pair at the resonant excitation energy where the effects of the optical alignment and the ptical orientation are observed.

Thus, the effects of the optical alignment and optical orientation of excitons are observed in core/shell CdSe/CdS NPLs. Our detailed studies of these effects and of the time-resolved PL decay allowed us to confirm that at the high energy edge of the spectrum the PL mostly comes from the exciton recombination. Further, we focus only on the effects of the optical alignment and optical orientation of excitons and develop a theoretical analysis for the excitons with large bright-dark splitting. We consider only the contribution from the horizontally lying NPLs ($\bm c \parallel z$, Figure \ref{fig:npl_geom}(b)) taking into account their in-plane random orientation.

\section{Theory}\label{sec:Theory}

We develop a theory of the optical alignment and optical orientation of excitons in the face-down NPLs oriented horizontally at the substrate as schematically shown in Figure \ref{fig:npl_geom}(b). The $\bm c$ axis of all NPLs under consideration is thus directed perpendicular to the substrate and along the $[001]$ crystallographic direction. Generally, the NPL edges can be directed either along $[100]$ and $[010]$ or along  $[110]$ and $[1\overline{1}0]$ directions~\cite{NPLanisotropy} and be of different length. The in-plane anisotropy of the studied NPLs is not large \cite{Feng2020}, however, is present. Therefore, we introduce two coordinate frames: the laboratory frame with axes $x,y,z$ and the frame related to the NPL with axes $X,Y$ along NPL edges and $Z$  axis directed along ${\bm c}$ as shown in Figure \ref{fig:theory}(a).  We consider the normal incidence of the exciting light to the sample with ${\bm k} \parallel z \parallel Z$ and reverse direction of the detected light (Figure \ref{fig:angle_avg_bright}), so that the vector of the light polarization ${\bm e} = (e_x,e_y,0)$ is always in the NPL plane. However, the $X,Y$ axes of the NPL frame might be rotated by angle $\alpha$ with respect to the $x,y$ axes of the laboratory frame. In the following consideration, the external magnetic field is applied in the Faraday geometry ${\bm B} \parallel {\bm k} \parallel \bm c$.

\subsection{Bright and dark exciton contributions to the PL polarization}\label{sec:Total polarization}

The band-edge exciton states in CdSe based NPLs comprise bright and dark excitons, for more details see SI, section S1. These excitons are formed from the electrons with the spin projection $s_Z = \pm 1/2$ on the NPL ${\bm c}$ axis and the heavy holes with the spin projection $j_Z= \pm 3/2$.
In the absence of the external magnetic field and any anisotropy-related splittings, the bright (A) and dark (F) excitons have two-fold degenerate  $| \pm1 \rangle$ states described by the wave functions $\Psi_{m_A}$ with projections $m_A=s_Z + j_Z= \pm 1$ and two-fold degenerate $| \pm2 \rangle$  states $\Psi_{m_F}$ described by the wave functions with projections $m_F=s_Z + j_Z= \pm 2$. Due to small perturbations caused by interactions with phonons or internal magnetic fields, the $| + 2 \rangle$ ($| - 2 \rangle$) states are coupled to the $| + 1 \rangle$ ($| -1  \rangle$) states and emit circularly polarized light \cite{Rodina_2016}.

In the absence of an external magnetic field the bright (dark) exciton states are split  by $\hbar \Omega_X$ ($\hbar \Omega_{FX}$) into linearly polarized dipoles $| X \rangle, | Y \rangle$ (and analogously composed $| FX \rangle, | FY \rangle$) and described by the wave functions
\begin{eqnarray}
\Psi_{X} = \frac{\Psi_{+1}+\Psi_{-1}}{\sqrt{2}} \, , \quad \Psi_{Y} = -i \frac{\Psi_{+1}-\Psi_{-1}}{\sqrt{2}} \, , \\
\Psi_{FX} = \frac{\Psi_{+2}+\Psi_{-2}}{\sqrt{2}} \, , \quad \Psi_{FY} = -i \frac{\Psi_{+2}-\Psi_{-2}}{\sqrt{2}} \, .
\end{eqnarray}
The anisotropic splitting $\hbar\Omega_{X}$ is associated with the long-range electron-hole exchange interaction in the presence of the anisotropy of the NPL shape in the plane \cite{Goupalov1998, Hu2018}.  The splitting between the dark exciton states $\hbar \Omega_{FX}$ has a different nature -- even without an in-plane anisotropy it can originate from the cubic-anisotropy contribution to the short-range electron-hole exchange interaction  $\sim~\sum_{\alpha=X,Y,Z} \sigma_{\alpha} J_{\alpha}^3$, where $\bm \sigma$ is a pseudo-vector composed of Pauli matrices and $\bm J$ is a pseudo-vector of matrices of the angular momentum $3/2$ \cite{IvchenkoPikus_book}. The fine structure of the band-edge exciton taking into account the anisotropic splitting in the absence of an external magnetic field is shown schematically in Figure S1(b).

\begin{figure*}[ht]
	\includegraphics[width=0.32\linewidth]{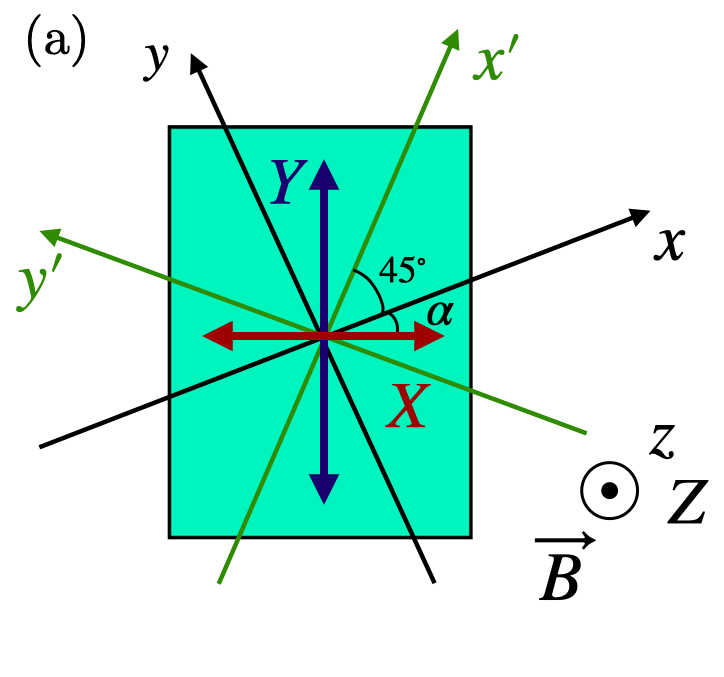}
	\includegraphics[width=0.59\linewidth]{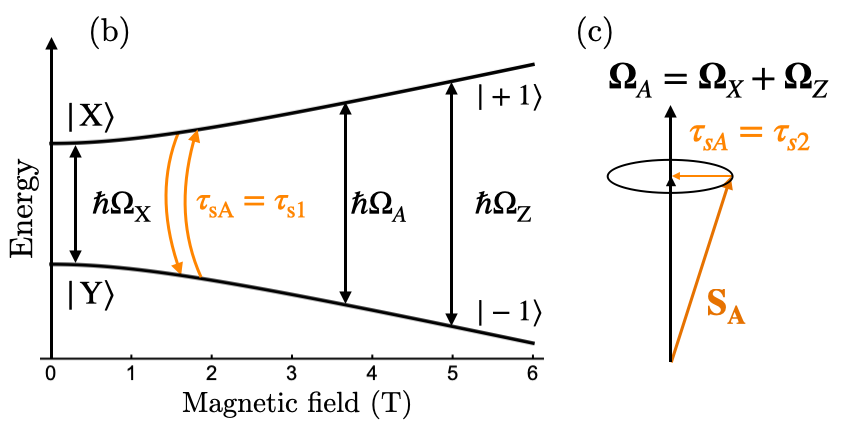}
	\caption{\label{fig:theory}  (a) A scheme of a single nanoplatelet with its own axes $X,Y$ and laboratory frame: $x, y, z$ and rotated by $45^{\circ}$ around $z$-axis $x', y'$, the magnetic field in the Faraday geometry. (b) Representation of the conversion of linearly polarized states $| X \rangle, |Y\rangle$ of the bright exciton, splitted by the energy $\hbar \Omega_X$, into circularly polarized components $| +1 \rangle, |-1\rangle$ in the case of $g_A>0$. A similar transformation takes place for dark exciton states $|F X \rangle, |FY\rangle, | +2\rangle, |-2\rangle$ and is shown in Figure S1(c). (c) Pseudospin $\bm S_A$ in the effective magnetic field with frequency $\bm \Omega_A$. Anisotropic relaxation times $\tau_{s1,s2}$ correspond to longitudinal (b) and transverse (c) relaxation, respectively. }
\end{figure*}

The Hamiltonian, taking into account  the electron-hole exchange terms and the Zeeman field-induced term for $\bm B \parallel \bm z$ for the exciton wave function in the basis $ \{\Psi_{+1}, \Psi_{-1}, \Psi_{+2}, \Psi_{-2}\}$, has the matrix form:

\begin{widetext}
	\begin{equation}\label{ham}
	\mathcal{H}_{AF} =\begin{pmatrix}
	\mathcal{H}_A &0\\
	0& \mathcal{H}_F\end{pmatrix} = \frac{\hbar}{2}\begin{pmatrix}
	\Omega_Z&  \Omega_X  &0&0\\
	\Omega_X & - \Omega_Z&0&0 \\
	0&0& -2\Delta E_{AF}/\hbar+ \Omega_{FZ}&  \Omega_{FX} \\
	0&0&\Omega_{FX}&-2\Delta E_{AF}/\hbar- \Omega_{FZ}
	\end{pmatrix}, 
	\end{equation}
\end{widetext}
where $\hbar\Omega_Z = g_A \mu_B B$, $\hbar\Omega_{FZ} = g_F \mu_B B$, $g_{A,F}$ -- bright, dark exciton $g$-factors, $\mu_B$ -- Bohr magneton. The Hamiltonian \eqref{ham} does not include any perturbations that can directly mix bright and dark exciton states. Therefore, the eigenstates of this Hamiltonian, $\Psi_A^\pm$ and $\Psi_F^\pm$, comprise only the linear combinations of the functions $\Psi_{\pm 1}$ and $\Psi_{\pm 2}$, respectively, with the energy eigenvalues $E_A^\pm(B)$ and $E_F^\pm(B)$: 
\begin{eqnarray}
E_A^\pm &=& \pm \frac{1}{2}\hbar  \Omega_A = \pm \frac{1}{2} \hbar\sqrt{\Omega_X^2+\Omega_Z^2} \, , \\ \nonumber
E_F^\pm &=&  - \Delta E_{AF} \pm \frac{1}{2} \hbar \Omega_F = \\
&=& - \Delta E_{AF} \pm \frac{1}{2} \hbar\sqrt{\Omega_{FX}^2+\Omega_{FZ}^2} \, . 
\end{eqnarray}
The exciton energy structure and its evolution in the magnetic field is shown in SI (section S1, Figure S1). The evolution of the bright exciton states in external magnetic field is shown in a larger scale in Figure \ref{fig:theory}(b).

The polarization of light emitted by excitons can be described using the spin density matrix formalism \cite{Dzhioev1997PRB, Blum_book}. The strong exchange interaction leads to a large splitting between the states of the bright and dark excitons $\Delta E_{AF}\sim1\div5$~meV, that is much larger than the inverse exciton lifetimes. It makes the states of the dark and bright excitons incoherent to each other. This, in turn, allows us to neglect non-diagonal block terms of the density matrix $\rho_{AF, m_A m_F}, \rho_{AF, m_F m_A} $ and  to write the block-diagonal density matrix for the four exciton states
\begin{equation}\label{matrix}
	\rho_{AF} = \begin{pmatrix}
		\rho_A & 0 \\
		0 & \rho_F
	\end{pmatrix}  = \begin{pmatrix}
\rho_{+1,+1} & \rho_{+1,-1} &0 &0 \\ 
\rho_{-1,+1} & \rho_{-1,-1} & 0 &0\\
	0 & 0 &  \rho_{+2,+2} & \rho_{+2,-2} \\
	0 & 0 &\rho_{-2,+2} & \rho_{-2,-2} 
\end{pmatrix},
\end{equation}
where the $2\times 2$ density matrices $\rho_{A}$ and $\rho_{F}$ characterize isolated two-level systems of bight and dark exciton states $\{ \Psi_{+1}, \Psi_{-1}\}$ and $\{ \Psi_{+2}, \Psi_{-2}\}$, respectively. For both two-level systems we can introduce bright and dark exciton pseudospin $\bm s_A$ and $\bm s_F$, respectively, and express the bright and dark density matrices as
\begin{equation}\label{key}
\rho_{A,F} = \frac{1}{2} N_{A,F} + \bm \sigma \cdot \bm s_{A,F} = \frac{N_{A,F}}{2} \left(1  + \bm \sigma \cdot \bm S_{A,F} \right) .
\end{equation}
Here $N_A =  (\rho_{+1,+1} + \rho_{-1,-1})$ and $ N_F=  (\rho_{+2,+2} + \rho_{-2,-2})$ are the bright and dark exciton populations and the average pseudospins are $\bm S_A = \bm s_A/N_A, \bm S_F = \bm s_F/N_F$ with $S_{A,F} = 1/2$. 

The bright exciton states with $m_A= \pm 1$ absorb and emit circular polarized light. The dark exciton states $m_F= \pm 2$  interact with light only due to the admixture of $m_A= \pm 1$  states and inherit their circular polarization selection rules. The direct absorption of the light by the dark exciton states can be safely neglected, however, their contribution to the PL is important at low temperatures due to their significant population. Accordingly, the total intensity  from the colloidal NPLs $I = I_A+I_F$ consists of the intensities of the bright exciton $I_A = \Gamma_A N_A $ and the dark exciton $I_F = \Gamma_F N_F$ (the recombination rates $\Gamma_{A,F}$ of both excitons are assumed to be purely radiative).

Then, the polarization of light emitted by excitons from a single horizontally oriented NPL with axes $X, Y$ can be presented as: 
	\begin{eqnarray}
	\label{key_all}
	P_{C} =i(e_Xe_Y^*-e_X^*e_Y) = {\cal A} P_{CA} + {\cal F} P_{CF} \, , \nonumber \\
	P_{L} = |e_X|^2-|e_Y|^2 ={\cal A} P_{LA} + {\cal F}P_{LF},  \\
	P_{L'} = e_Xe_Y^*+e_X^*e_Y = {\cal A} P_{L'A} + {\cal F} P_{L'F}. \nonumber
\end{eqnarray}
Here $e_X$ and $e_Y$ are the projections of the emitted light polarization vector ${\bm e}$ on the $X,Y$ axes, ${\cal A} = I_A/(I_A+I_F)$ and ${\cal F} = I_F/(I_A+I_F)$ characterize the contributions from the bright and dark exciton correspondingly. The partial light polarizations are related to the density matrix and  averaged pseudospins components as 
\begin{widetext}
\begin{eqnarray}
	\label{key1}
	P_{CA}  = \frac{\rho_{+1,+1} - \rho_{-1,-1}}{\rho_{+1,+1} + \rho_{-1,-1}}  = 2 S_{AZ}\, ,  \quad P_{CF} =  \frac{\rho_{+2,+2} - \rho_{-2,-2}}{\rho_{+2,+2} + \rho_{-2,-2}}   = 2 S_{FZ} \, ,   \\
	P_{LA} =  -\frac{\rho_{+1,-1} + \rho_{-1,+1}}{\rho_{+1,+1} + \rho_{-1,-1}}  = 2 S_{AX}\, , \quad P_{LF} =  -\frac{\rho_{+2,-2} + \rho_{-2,+2}}{\rho_{+2,+2} + \rho_{-2,-2}}  = 2 S_{FX}\, ,\nonumber \\
	P_{L'A}  = i\frac{\rho_{+1,-1} - \rho_{-1,+1}}{\rho_{+1,+1} + \rho_{-1,-1}}  = 2 S_{AY} \, ,  \quad P_{L'F} =  i\frac{\rho_{+2,-2} - \rho_{-2,+2}}{\rho_{+2,+2} + \rho_{-2,-2}}  = 2 S_{FY}\, . \nonumber 
\end{eqnarray}
\end{widetext}

In the laboratory frame the registered polarization depends on the orientation of the NPL rotated around the laboratory axis $z$ by an angle $\alpha$ (Figure \ref{fig:theory}(a)). Meanwhile, the total intensity  $I$ does not depend on the angle $\alpha$. Therefore, the linear polarizations registered in the laboratory frame depend on the rotation angle of the single NPL as follows
\begin{gather}\label{angle}
P_{l}(\alpha) = P_{L} \cos (2\alpha) + P_{L'} \sin (2\alpha),  \nonumber \\
P_{l'}(\alpha) = -P_{L} \sin (2\alpha) + P_{L'} \cos (2\alpha), \\
 P_{c} = P_{C} \nonumber .
\end{gather}

Thus, to describe the effects of the optical orientation and alignment of the excitons we need to find first the components of the bright and the dark exciton pseudospins in the external magnetic field. 

\subsection{Pseudospin components in the magnetic field in the Faraday geometry}\label{sec:Pseudospin} 

The dynamics of the density matrix $\rho_{AF}$  is defined by the equation:
\begin{equation}\label{key_rho}
\frac{\partial \rho_{AF}}{\partial t} = \frac{1}{i \hbar} [ \mathcal{H}_{AF}, \rho_{AF} ],
\end{equation}
allowing us to obtain two separate equations for the bright and dark exciton pseudospin $\bm s_{A,F}$:
\begin{gather}\label{syst_simplified}
\frac{d\bm s_A}{dt} + {\bm s_A} \times {\bm \Omega_A} = 0,  \ \ 
\frac{d\bm s_F}{dt}+ \bm s_F \times \bm \Omega_F = 0 .
\end{gather}
Each pseudospin rotates in its own effective magnetic field with a frequency $\bm \Omega_{A,F} = \bm \Omega_{X,FX} +\bm \Omega_{Z,FZ}$ that include both the external magnetic field and the field directed along $X$ pseudospin component that causes the anisotropic splitting.

The interaction between the bright and the dark exciton pseudospins is realized by their mutual pumping during the relaxation between them, which can be accounted for in the framework of the kinetic equations. The pseudospins dynamics taking into account generation, relaxation and recombination is described by a system of kinetic equations, phenomenologically written as follows
\begin{gather}
\frac{d\bm s_A}{dt} +\frac{{\bm s_A}}{T_A} + {\bm s_A} \times {\bm \Omega_A} = \frac{\bm s_A^0}{\tau_A} \label{sa}, \\
\frac{d\bm s_F}{dt}+\frac{\bm s_F}{T_F} + \bm s_F \times \bm \Omega_F = \frac{\bm s_F^0}{\tau_F} \ . \label{sf}
\end{gather}
Here the right-hand  terms describe the generation of bright and dark pseudospins both due to the initial pumping by the polarized light and due to the mutual relaxation between bright and dark excitons and will be discussed in section \ref{sec:Generation}. $T_{A,F}$ are pseudospin lifetimes $1/T_{A,F} = 1/\tau_{A,F}+1/\tau_{sA,sF}$, where $\tau_{A,F}$ are the exciton lifetimes specified in SI, Eq. \eqref{tau} and $\tau_{sA,sF}$ are the spin relaxation times. Note, that the kinetic Eqs. \eqref{sa}, \eqref{sf} should be supplemented by the system of rate equations for the bright and the dark exciton populations $N_A$ and $N_F$ given in SI, Eq. \eqref{populations_cw}.

In general case, the exciton spin relaxation times and therefore the exciton pseudospin lifetimes $T_{A,F}$ have to be considered as the tensors of the second rank. Depending on the relaxation mechanism, the main axis of these tensors and the main values may depend both on the value and on the direction of the effective fields  ${\bm \Omega}_{A,F}$. The consideration of the particular mechanism is beyond the scope of the present paper.

For the bright exciton we consider two possibilities: (i) isotropic spin relaxation times allowing to consider $T_A$ as a scalar; (ii) anisotropic spin relaxation times assuming that $T_A$ has two independent values $T_1$ and $T_2$ related to the longitudinal relaxation time $\tau_{s,1}$ accompanied by the energy relaxation between the eigenstates $\Psi_A^\pm$ and the dephasing time $\tau_{s,2}$ caused by the pseudospin rotation around the effective field ${\bm \Omega}_A$ (see Figure \ref{fig:theory}(b),(c)).
For the dark exciton, we restrict the consideration by assuming the isotropic spin relaxation times, which allows us to consider $T_F$ as a scalar. As it will be shown later, this is sufficient to describe the effects under study. 

In the following, we consider only steady state solutions of Eqs. (\ref{sa}, \ref{sf}) under cw excitation. In this case, the averaged pseudospins  $\bm S_{A}=\bm s_{A}/N_A^0$ and $\bm S_{F}=\bm s_{F}/N_F^0$, where  $N_{A,F}^0$ are the steady state solutions of the rate Eqs. (\ref{matrixA}, \ref{populations_cw}), satisfy the same system of kinetic Eqs. (\ref{sa}, \ref{sf}). Then, in the case of the isotropic spin relaxation times and under the additional restrictions 
$\Omega_F T_F \gg 1$ and $\Omega_A T_A \gg 1$, the steady state solutions  can be written as
\begin{equation}\label{limit}
{\bm S}_A = \frac{T_A}{\tau_A} \frac{({\bm S}_A^0{\bm \Omega}_A ){\bm \Omega}_A}{{\Omega}_A^2} \, , \quad {\bm S}_F = \frac{T_F}{\tau_F} \frac{({\bm S}_F^0{\bm \Omega}_F ){\bm \Omega}_F}{{\Omega}_F^2} \, .
\end{equation} 
One can see, that under the considered restriction, allowing many pseudospin rotations around the effective field during its lifetime, the steady state pseudospin is always directed along the effective field \cite{Kusrayev2008, Dzhioev1997PRB}. It is easy to demonstrate that the same property remains even with the consideration of the spin relaxation anisotropy. Applied magnetic field in the Faraday geometry converts linearly polarized dipoles $|X\rangle, |Y\rangle$ into circular components $|\pm 1>$ (Figure \ref{fig:theory}(b)). This restores the circular polarization of PL when $\Omega_Z T_A \gg 1$.  However, the $S_Y$ component is always vanishing in the considered geometry and the $S_Z$ component is vanishing in zero magnetic field. For this reason, the above restrictions do not allow one to describe the effects of the rotation of the linear polarization plane and the optical orientation, which are observed in the experiment. We assume that the main effects come from the bright exciton recombination and, therefore, consider further the steady state  solutions of Eqs. (\ref{sa}, \ref{sf}) for ${\bm S}_A$ assuming arbitrary relations between the main values of the spin lifetime tensor $T_A$ and $\Omega_A$ for the cases (i), (ii).  

In the case of isotropic relaxation time (case (i)), the general solution for the pseudospin $\bm S_A$ steady-state components in a certain magnetic field in the Faraday geometry reads:
\begin{widetext}
	\begin{gather}\label{PL} \nonumber
	S_{AX} = \frac{T_A (S_{AX}^0(1+ \Omega_X^2 T_A^2)-\Omega_Z T_A S_{AY}^0+\Omega_Z\Omega_X T_A^2 S_{AZ}^0) }{\tau_A(1+\Omega_A^2 T_A^2)}, \\
	S_{AY} = \frac{T_A(S_{AY}^0+\Omega_Z T_A S_{AX}^0-\Omega_XT_A S_{AZ}^0)}{\tau_A(1+\Omega_A^2T_A^2)} ,\\
	\nonumber S_{AZ} = \frac{T_A(S_{AZ}^0(1+\Omega_Z^2 T_A^2)+\Omega_XT_A S_{AY}^0 +\Omega_X\Omega_Z T_A^2 S_{AX}^0)}{\tau_A(1+\Omega_A^2T_A^2)}.
	\end{gather}
\end{widetext}
The solutions for the case (ii) taking into account the anisotropy of the relaxation times are given in the SI, section S4.

\subsection{Pseudospin generation}
\label{sec:Generation}

We consider now the generation terms at the right-hand side of  the kinetic eqs (\ref{sa}, \ref{sf}).
We characterize the polarization of the exciting light by two Stokes parameters
\begin{gather}
\label{key_pump}
P_{c}^0 = i(e_x^0e_y^{0*}-e_x^{0*}e_y^0) \, , 
P_{l}^0 = |e_x^0|^2-|e_y^0|^2 \, , \\
P_{l'}^0 = e_x^0e_y^{0*}+e_x^{0*}e_y^0 \, , \nonumber
\end{gather}
where $e_x^0,e_y^0$ are projections of the light polarization vector ${\bm e}_0$ on the laboratory axes $x,y$. 
For the  resonant or close to the resonant excitation conditions we neglect the small oscillator strength of the dark exciton and assume that only $| \pm 1 \rangle$ states are excited by the laser.   
At low temperature, when there are no transitions from the dark exciton to the  bright one, the generation term for the bright exciton pseudospin $\bm S_A^0 = \bm S_0$ and $\bm s_A^0 = N_A^0\bm S_A^0$. For the NPL with axes $X, Y$, rotated about the laboratory axis $z$ by an angle $\alpha$, it is related to the exciting light polarization as
\begin{gather}\label{pump1}
S_{X}^0 = \frac{\gamma_l}{2} P_l^0 \cos(2\alpha), \nonumber \\
S_{Y}^0 = \frac{\gamma_l}{2} P_l^0 \sin(2\alpha),  \\
S_{Z}^0 = \gamma_c P_c^0  \, . \nonumber
\end{gather}
Here the parameters $\gamma_l \le 1$ and $\gamma_c \le 1$ are introduced to account for the possible loss of the pumped polarization during the relaxation in the case when the excitation is not exactly resonant. Hereafter, by $P_l^0 $ and $P_c^0$ we mean $\gamma_l P_l^0 $ and $\gamma_c  P_c^0$, respectively, thus taking into account the loss of the polarization for the excitons created in lying NPLs.  Moreover, analyzing the absolute values of the polarizations, one has to take into account the loss of the polarization caused by the contribution to the recombination coming from the vertically oriented NPLs. These losses can be also included in the factors $\gamma_l$ and $\gamma_c$.

We assume that the generation terms $\bm S_F^0$ and $\bm s_F^0=N_F^0 \bm S_F^0$  for the dark exciton do not contain the contribution from the laser excitation and originate totally from the relaxation of the polarized population from bright to dark exciton. In the case of a non-resonant excitation, the dark exciton state population $N_F^0$ can be created via the relaxation from the initially excited states and not from the band-edge bright exciton. However, in this case, the polarization will be lost during the relaxation process. We  assume that the perturbation matrix elements responsible for the relaxation from bright to dark exciton states admix the exciton states  $| +1\rangle$ to $| +2\rangle$ and  $| -1\rangle$ to $| -2\rangle$ with the same probabilities as shown in Figure S1(c). Therefore, the circular polarization is preserved during excitation transfer and $ S_{FZ}^0 = S_{AZ}$. As for the pseudospin components associated with the linear polarization, they can be lost due to the  phases of the wave functions. Below we assume that $S_{FY}^0 = 0$ and consider two extreme cases for the $S_{FX}^0$ component: (a) $S_{FX}^0 = 0$ and (b) $ S_{FX}^0 = S_{AX}$ due to the transfer of the liner polarization from the anisotropically split   $| X\rangle$ to $| FX\rangle$ and  $| Y\rangle$ to $| FY\rangle$ as shown in Figure S1(b). The case (b) can be realized, for example, if the coupling between bright and dark states is caused by the anisotropic internal magnetic field directed predominantly along $X$ axis. Consideration of the microscopic mechanisms responsible for this field or other total or partial transfer of the linear polarization from bright to dark exciton is beyond the scope of this paper. 

As the temperature increases, the generation terms are modified due to the acceleration  of the bright-to-dark and temperature activation of the dark-to-bright exciton relaxations:
\begin{gather}\label{SA0}
\bm S_A^0 = \left(1- f \right) {\bm S_0}+ f {\bm S_F^*}\, , 
\end{gather}
where $f = (1-\Gamma_A\tau_A)(1-\Gamma_F\tau_F)$.  The derivation of this expression is given in SI.  From hereafter, we keep only equations in low temperature regime in the main text.




The transfered terms $\bm S_F^*$ and $\bm s_F^* = N_F^0 \bm S_F^*$ also depend on the relaxation conditions between the bright and the dark excitons. We assume that $S_{FZ}^* =S_{FZ}$ and $S_{FY}^* = 0$ and for the two cases under consideration (a) $S_{FX}^* = 0$ or (b) $ S_{FX}^* = S_{FX}$.

\subsection{Bright exciton contribution to the PL polarization from the ensemble of NPLs}\label{sec:Orientation}

Recall that NPLs are randomly oriented in the plane of the substrate. For this reason, some contributions to the  polarization from a single NPLs, such as the conversion of the linear to circular polarization,  disappear upon averaging over the ensemble. Since the total intensity does not depend on the in-plane orientation of the NPL, only the averaging of  the polarizations coming from each NPL over the angle $\alpha$ is needed. The circular polarization is not affected by the angular dependence and the optical orientation  effect comes from the $S_{AZ}$ component generated by $P_c^0$. As for the linear polarized components, the nonvanishing contributions coming from the initial $P_l^0$ excitation and given by Eqs. \eqref{angle} and \eqref{pump1} for the bright exciton in the case (i) at low-temperature regime are the following contributions (see also orange arrows in Figure \ref{fig:angle_avg_bright}):
\begin{widetext}
\begin{gather}\label{angle}
P_{lA}^l(\alpha) = \frac{T_A P_{l}^0
}{\tau_A(1+\Omega_A^2 T_A^2)}\left( (1+\Omega_X^2T_A^2)\cos^2 (2\alpha) + \sin^2 (2\alpha)  \right), \\
P_{l'}^l(\alpha) =  \frac{T_A P_{l}^0\Omega_Z T_A}{\tau_A(1+\Omega_A^2 T_A^2)}\left(\sin^2 (2\alpha) +\cos^2 (2\alpha) \right).\nonumber \end{gather}
\end{widetext}
After averaging we get results for three Stokes parameters in the laboratory frame:
\begin{gather}\label{coni}
P_{lA}^l= \frac{T_A}{2\tau_A}\frac{P_{l}^0(2+\Omega_X^2 T_A^2) }{(1+\Omega_A^2T_A^2)}, \nonumber \\
P_{l'A}^l = \frac{T_A}{\tau_A}\frac{T_A\Omega_{Z} P_l^0}{(1+\Omega_A^2 T_A^2 )}, \\ 
\nonumber P_{cA}^c = \frac{T_A}{\tau_A}\frac{P_c^0(1+\Omega_{Z}^2 T_A^2)}{(1+\Omega_A^2 T_A^2)}.
\end{gather}
The conversion of the circular polarization to the linear and vice versa are vanishing after averaging.
The expressions for the effects alter with the consideration of the relaxation time anisotropy in the case (ii), and takes form:
\begin{gather}
P_{lA}^l = \frac{P_l^0}{2\tau_A}\left( \frac{T_2 (2+\Omega_X^2 T_2^2) }{(1+\Omega_A^2T_2^2)} + (T_1-T_2)\frac{\Omega_X^2}{\Omega_A^2}\right), \nonumber \\  \label{anis12}
P_{l'A}^l = \frac{P_l^0}{\tau_A}\frac{T_2^2\Omega_Z}{1+\Omega_A^2T_2^2} \, , \\ \nonumber
P_{cA}^c = \frac{P_c^0}{\tau_A}\left(\frac{T_2(1+\Omega_{Z}^2 T_2^2)}{(1+\Omega_A^2T_2^2)} + (T_1-T_2)\frac{\Omega_Z^2}{\Omega_A^2}\right).
\end{gather}
Eqs. \eqref{anis12} are transformed into Eqs. \eqref{coni} when $T_1=T_2=T_A$.

\subsection{Dark exciton contribution to the ensemble polarization}\label{sec:Dark1}

Let us consider two extreme cases of the relaxation between the bright and dark exciton states, described in the section \ref{sec:Generation}. All possible component transformations for the transfer of linear and circular components are shown in Figure S4. The detected contributions to the optical alignment and to the rotation of the linear polarization plane are highlighted on the right in orange and green, respectively. If only $S_z$ component of the pseudospin is conserved during excitation transfer from the bright exciton to the dark one (solid red and violet arrows in Figure S4), only two contributions each from the dark exciton occur in $P_l^l$ and $P_{l'}^l$ at low temperature.  If there is a generation of the $S_{FX}^0$ as well (dashed red arrows)  in Figure S4), both effects  include two additional contributions more (dashed violet arrows each in Figure S4).

\begin{figure*}[ht]
	\includegraphics[width=0.8\linewidth]{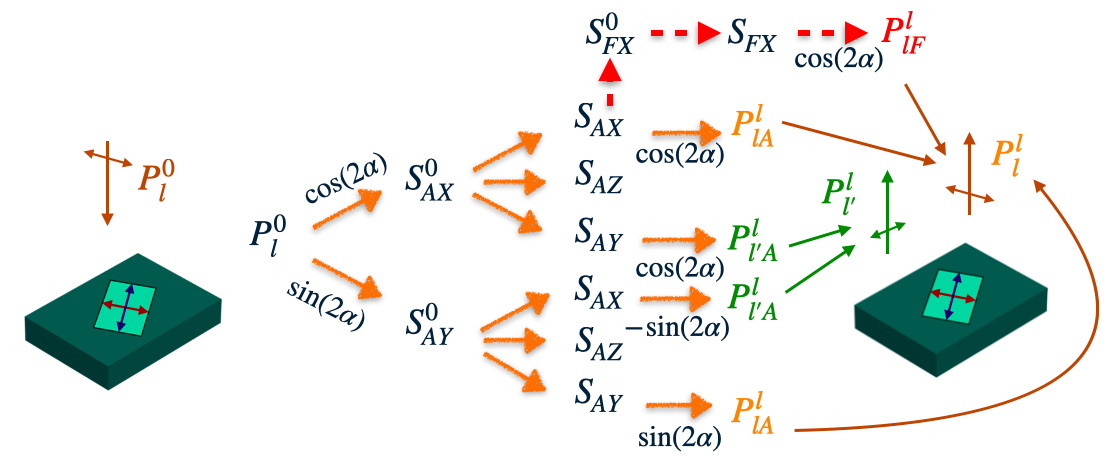}
	\caption{\label{fig:angle_avg_bright} Scheme of the linear polarization $P_l^0$ conversion in the bright exciton into detectable effects of optical alignment $P_l^l$ (marked in orange) and rotation of the linear polarization plane $P_{l'}^l$ (marked in green) for the NPL oriented at the angle $\alpha$ (Figure \ref{fig:theory}(a)). The main contribution to the optical alignment from the dark exciton is shown on the top with red arrows (See in more details in SI, section S3).}
\end{figure*}

Here we give the solutions taking into account the anisotropy of the relaxation time in the bright exciton for the two notable contributions. Answers for all the contributions to the measured polarization from the dark exciton at low temperature without taking into account the anisotropy of the relaxation time in the bright one are given in the SI, section S3. 

The effect of the optical orientation in the case of the only transferred $S_Z$ to $S_{FZ}$ consist in one term for $P_{c}^c$:
\begin{equation}\label{pcF}
P_{cF}^c = - \frac{T_F P_c^0}{\tau_F}\frac{(1+\Omega_{FZ}^2T_F^2)}{(1+\Omega_F^2T_F^2)} \frac{(T_2 \Omega_X^2+T_1\Omega_Z^2 (1+T_2^2\Omega_A^2) }{\tau_A \Omega_A^2(1+T_2^2\Omega_A^2)} \, .
\end{equation}
 One can see, that in zero external magnetic field $P_{cF}^c$ is vanishing in the limit case $\Omega_{FX}T_F \gg 1$, which as we will see below is well satisfied for the dark exciton in our NPLs. 
 
Activation of the pseudospin linear component transfer enables the most significant in magnitude contribution from the dark exciton to the optical alignment by direct transfer of $S_{X}$ to $S_{FX}$ (shown by red dashed arrows in Figure \ref{fig:angle_avg_bright}):
\begin{equation}\label{AlignF}
P_{lF}^l = \frac{T_F P_l^0}{2\tau_F} \frac{(1+\Omega_{FX}^2T_F^2)}{(1+\Omega_F^2T_F^2)} \frac{T_2 \Omega_Z^2+T_1 \Omega_X^2(1+T_2^2\Omega_A^2)}{\tau_A \Omega_A^2(1+T_2^2\Omega_A^2)}.
\end{equation}
Other nonvanishing contributions are given in SI,  see S3.

It is important to note, that all effects of the conversion from linear to circular polarization in bright or dark exciton are vanishing upon averaging over randomly oriented NPL ensemble. All nonvanishing contributions of the dark exciton state to the optical alignment effect are shown in  SI, Figure S4. In addition to the random orientation of the NPLs in ensemble, the exciton parameters in the NPLs, such as the anisotropic exciton splittings  $\Omega_X$ and $\Omega_{XF}$, can be characterized by some dispersion in the ensemble. To account for it, an additional averaging with the distribution function can be carried out, as it has been done, for example, in Ref. \cite{Nestoklon2018} for the ensemble of perovskite nanocrystals.

To sum up, here, in this section, we have given the theoretical blocks necessary to describe the polarization effects in the ensemble of colloidal nanoplatelets. Below comes the analysis of experimental data by means of the developed theory in order to determine quantitative ranges for the exciton parameters of the ensemble of NPLs.

\section{Modeling of experimental data and discussion}\label{sec:Discussion}

\subsection{\label{sec:Field analysis}Analysis of the magnetic field dependences of the Stokes parameters}

In this subsection, we analyze the experimental magnetic-field dependences of three Stokes parameters, $P_l^l$, $P_{l'}^l$, and $P_c^c$, shown in Figure \ref{cw}(b) and \ref{cw}(d) in for the low temperature regime. We start from the contributions coming from the bright exciton resulting in the broad contours in all three Stokes parameters in Figure \ref{cw}(b).
	
Regardless of the amplitudes, we can write down three conditions on the field dependences in a fixed magnetic field. In the field of $B_{1/2} = 3$ T the half-maximum (HWHM) of the optical alignment effect is reached. The pseudospin lifetime can be different for longitudinal and transverse relaxation, and can also depend on the magnitude of the external magnetic field as was discussed in Section \ref{sec:Theory}. To avoid an excessive number of undetermined parameters, we analyze two extreme cases, neglecting either the field dependence of the spin lifetime or anisotropy of the relaxation time.

Here in the main text we consider the cases (i) without  and (ii)  with account of the pseudospin relaxation anisotropy but  assume that $T_A, T_1,T_2$ do not depend on the magnetic field. In the SI (see S4), we consider the case (iii) assuming $T_1= T_2 = T_A$ but taking into account the  dependence of the $T_A$  on the magnetic field. For clarity, we introduce the variables $X_A = \Omega_X T_A = \Omega_X T_1$ and $Z_A = \Omega_Z (B_{1/2}) T_A = \Omega_Z (B_{1/2}) T_1$.

Other two conditions are the ratio of the linear polarizations $P_{l'A}^l/{P_{lA}^l}$ in a $B_{1/2}$ field and the recovery of the initial optical orientation in the same field $P_{cA}^c(0)/{P_{cA}^c(B_{1/2})}$. In the case (i), without taking into account the anisotropy of the relaxation time and its magnetic field dependence, these conditions take the form:
\begin{gather}\label{Conditions}
\frac{P_{lA}^l(B_{1/2})}{P_{lA}^l(0)}= \frac{1+X_A^2}{1+X_A^2+Z_A^2}= \frac{1}{2}\ , \\  \frac{P_{l'A}^l}{P_{lA}^l}(B_{1/2}) = \frac{2Z_A}{2+X_A^2}, \nonumber \\ \frac{P_{cA}^c(0)}{P_{cA}^c(B_{1/2})} = \frac{1+X_A^2+Z_A^2}{(1+X_A^2)(1+Z_A^2)}.
\end{gather}
All three conditions for magnetic field dependences for the two variables $X_A, Z_A$ generally cannot be satisfied simultaneously, as can be seen in blue curves in Figure S5. Figure S5(a) corresponds to the first of conditions \eqref{Conditions} for the optical alignment (HWHM condition), Figures S5(b) and S5(c) correspond to two other ratios plotted for the parameters satisfying the first HWHM condition. The horizontal dashed lines indicate the experimental values $P_{l'}^l(B_{1/2})/P_{l}^l(B_{1/2}) = 0.62$ and $P_{c}^c(0)/P_{c}^c(B_{1/2}) = 0.5$ extracted from Figure \ref{cw}(b).

Accounting for relaxation times anisotropy (the case (ii)) leads to the appearance of the additional parameter characterising the anisotropy $t_{12} = T_1/T_2$, where $T_1, T_2$ are related to relaxation times $\tau_{s,\parallel}$ and $\tau_{s,\perp}$ respectively (Figures \ref{fig:theory}(b), \ref{fig:theory}(c)). Analogous conditions can be obtained from Eqs. \eqref{anis12} and given in SI, see Eqs. \eqref{cond_anis}.

With these conditions in the case (ii) we obtain fixed parameter bindings $X_A =\Omega_X T_1 = 1$, $Z_A = \Omega_Z T_1 = 1.7$ and the pseudospin lifetime anisotropy $t_{12} = T_1/T_2 = 2$ (orange curves in Figure S5). These bindings, however, give the infinite number of parameter sets $(T_1,\  g_A = \hbar Z_A/\mu_B B_{1/2}, \ \hbar \Omega_X = \hbar X_A/T_1)$ which describe a broad part of all three magnetic field dependences shown in Figure \ref{cw}(b) for the detection close to the resonance.

Alternative set of the fixed parameters can be found for the case (iii) by assuming the magnetic field dependence of the spin lifetimes $T_1=T_2=T_A(B)$ as discussed in SI, see S4 and green curves in Figure S4. The obtained parameters for $X_A=\Omega_X T_A= 2.08$, $X_A(B_{1/2})=\Omega_X T_A(B_{1/2})= 1.5$ and  $Z_A(B_{1/2}) = \Omega_Z T_A(B_{1/2}) =1.22 $ are slightly different  from the previously found.  Next, similar procedures is done for the magnetic-field dependences shown in Figure \ref{cw}(c) for the slightly non-resonant detection energy. The resulting parameters of the bright exciton are close to the parameters found for the resonance detection (see SI, section S4). 

We next analyze the narrow part in the $P_l^l(B)$ dependence which can be seen in Figure \ref{cw}(b) and  is absent in Figure \ref{cw}(d). We focus here first only on its magnetic-field dependence, however its amplitude is zero in the case (a) and nonzero only in the case (b), when the linear polarization is transfered from  bright to dark exciton as shown in Figure S5 by the red dashed arrow. 
The HWHM of the narrow contour is reached in the field of $B_{F1/2} = 60 \ \rm mT$. In the main contribution to $P_{lF}^l$ of Eq. \eqref{AlignF} the terms  associated with the bright exciton in such a small field  are still close to those at zero field. Therefore, the condition for the HWHM of the narrow contour can be obtained as:
\begin{equation}\label{Plldark}
	\frac{P_{lF}^l(B_{F1/2})}{P_{lF}^l(0)} = \frac{(1+\Omega_{FX}^2T_F^2)}{(1+\Omega_F^2T_F^2)} = \frac{1}{2}\, . 
\end{equation}
In the limit case $\Omega_{FX}T_F \gg 1$  this condition results in $\Omega_{FZ} (B_{F1/2}) = \Omega_{FX}$. We will see below that this condition is well satisfied for the dark exciton. 

Thus, we have specified the fixed parameter bindings. Nevertheless, the goal of our work is to determine some specific ranges of possible exciton parameters. To extract the parameters of interest $\Omega_{X,FX}$ and $g_{A,F}$, let us analyze the amplitudes of the effects at low temperature.

\subsection{Amplitude analysis }\label{sec:Amplitude analysis}

 In this subsection we consider the amplitudes (the values at zero magnetic field) of the effects of optical alignment, $P_{l}^l(0)$,  and  optical orientation, $P_{c}^c(0)$, at low temperature. As we mentioned above, the amplitudes can be affected by the depolarization factor coming from the recombination of the excitons in the vertically oriented (edge-up) NPLs as well as by the initial loss of the polarization in the case of the non-resonant excitation. Both effects will be taken into account by allowing the initial polarizations $P_l^0,P_c^0 \le 1$. Next, we recall that there are two contributions to the total polarizations coming from the bright and dark excitons and weighted by the factors 
 ${\cal A} = I_A/(I_A+I_F) = \Gamma_A \tau_A$ and  ${\cal F} = I_F/(I_A+I_F) = 1-{\cal A} = \gamma_0 \tau_A$ at low temperatures. 
 Then, the bright  exciton contributions  to  the optical alignment and orientation effects are given by:
 	\begin{eqnarray}\label{ampA}
{\cal A} P_{lA}^l(0) &=&P_{l}^0 \frac{\Gamma_{A}}{2}\left( T_1+ \frac{T_2}{1+\Omega_{XA}^2T_2^2} \right) \, , \\
{\cal A} P_{cA}^c(0) &=&  P_c^0\frac{ \Gamma_A T_2}{1+\Omega_{XA}^2T_2^2} \, . 
 \nonumber
\end{eqnarray}
 The dark  exciton contribution   to  the amplitude of  optical alignment effect is given by (see Eq. \eqref{AlignF}):
 \begin{equation}
\label{ampF}
{\cal F} P_{lF}^l(0) = P_l^0\frac{T_1 T_F  \gamma_0  }{2\tau_F}  \, ,
\end{equation}
while its contribution to  the amplitude of the optical orientation effect is vanishing in the case $\Omega_{FX}T_F \gg 1$ (according to Eq. \eqref{pcF}).

At this stage we recall the exciton parameters defined from the PL decay data: $\Gamma_A=0.8$ ns$^{-1}$, $\gamma_0=0.31$ ns$^{-1}$ and $\tau_F =\Gamma_F^{-1}=250$ ns. We can use now the experimental amplitude 0.1 of the broad contour of the linear alignment effect at low temperature (see Figure \ref{cw}(a)) as the bright  exciton contribution and  determine the value of the $P_l^0 T_1$ product from Eq. \eqref{ampA}. We use the previously determined sets of the parameters bindings and obtain in the case (ii) $P_l^0 T_1 \approx 0.18$ ns  with $X_A = \Omega_X T_1 = 1$, $T_2 = T_1/2$, and alternatively in the case (iii) $P_l^0 T_1 \approx 0.21$ ns  with $X_{A} = \Omega_X T_1(0) = 2.08$ and $T_2 = T_1$. As $P_l^0 \le 1$, this gives us the lower limit for the values of $T_1$. The upper limit $T_1 \le \tau_A = 0.9$ ns comes from the bright exciton lifetime. 

 For the range of the bright exciton pseudospin lifetime $0.2{\rm \ ns}  \leq T_1\leq 0.9 {\rm \ ns}$, obtained in the case (a), we get the intervals for anisotropic splitting of states and $g$-factor of the bright exciton:  $\hbar \Omega_X \in [0.9, 5.9] \rm \ \mu eV$ and $g_A = 0.034 \pm 0.015$ (see Table S1 in the SI). Corresponding sets of the parameters allow us to describe the dependences on the magnetic field in the Faraday geometry of the optical alignment, the rotation of the linear polarization plane and the optical orientation, shown in Figure \ref{cw}(b). We used $P_l^0= 0.75$ and $P_c^0=0.47$. The alternative set in the case (iii) gives us  $\hbar \Omega_X \in [1.5, 6.2] \rm \ \mu eV$ and $g_A = 0.02 \pm 0.011$ (see Table S2 in the SI). Numerical analysis of amplitudes is presented in the S3. The data analysis of the amplitudes for the data from Figure \ref{cw}(d) is presented in S3.

Next, we turn to the parameters of the dark exciton. We assume that the narrow contour in Figure \ref{cw}(b) originates from the dark exciton contribution to the optical alignment effect. For its amplitude, we can estimate the lifetime of the dark exciton spin as $T_F=125$~ns in the case (a) or $T_F=108$~ns in the case (b). These values imply the dark exciton spin relaxation times $\tau_{sF} = 252$~ns or $\tau_{sF} = 189$~ns, respectively, comparable with the dark exciton lifetime $\tau_F=250$~ns.

The longitudinal $g$-factors of the bright and dark exciton in colloidal NPLs are determined by the expression:
\begin{gather}\label{key}
g_{A} = - g_e -3 g_h, \nonumber \\
g_{F} = g_e -3 g_h
\end{gather}
where $g_{e,h}$ are the electron and hole $g$-factors. The value of $g_e = 1.67$ for electron is obtained by spin-flip Raman scattering spectroscopy (Fig.~\ref{SFRS}{b}). However, the hole $g$-factor is unknown for the hole in the studied NPLs. In CdSe NPLs with thick shell in Ref. \cite{Shornikova2018nl} the value of $g_h = -0.4$ was obtained in low magnetic fields. By knowing the $g$-factor of the bright exciton $g_A$, we determine the range of the $g_h$ values comparable with the value from Ref. \cite{Shornikova2018nl} and thereby also determine the range of the values  $g_F = 3.36 \pm 0.015$  for the dark exciton $g$-factor.

 And at last, using the relation $\Omega_{FX} = \Omega_{FZ}(B_{F1/2})$  and determined $g_F$, we estimate the anisotropic splitting of the dark exciton in zero magnetic field as $\hbar \Omega_{FX} = (11.65 \pm 0.05 )\ \rm \mu eV$. Thus, for the determined values of $T_F$ we obtain $\Omega_{FX}T_F \sim 2000$ to justify the approximation $\Omega_{FX}T_F \gg 1$ used. The fact that the splitting between the dark exciton states in zero magnetic field turned out to be larger than the anisotropic splitting between the bright exciton states does not contradict the conditions of the problem. These splittings may have a different nature. The splitting between the states of the bright exciton is driven by the anisotropy of the NPLs in the plane which is not large for the studied NPLs. For the dark exciton, the cubic by $J_\alpha$ term of the Hamiltonian also leads to the splitting of its states. 
 
 Thus, we have determined all parameters of the bright and dark exciton fine structure with a good accuracy. The largest uncertainty remain about the anisotropic splitting between spin sublevels of the bright exciton and the relaxation time between them. For better access to these parameters as well as for the analysis of the temperature dependences of the amplitudes of the optical alignment (see Figure \ref{cw}(c)) we plan the future time-resolved studies of the optical alignment and optical orientation effects in NPLs.

\section{Conclusions}
\label{sec:conclusions}


We present the first observation of exciton optical alignment and optical orientation effects in the ensemble of core/shell CdSe/CdS colloidal NPLs and develop a model describing the dependences of these effects on the magnetic field in the Faraday geometry. The presence of a radiative recombination from the dark exciton state leads to the  appearance of two contours of the optical alignment in a longitudinal magnetic field -- the broad one and the narrow one. The transition of the  optical alignment from the bright to the dark exciton takes place only under the strongly resonant conditions (when the difference between detection and excitation energies is about  $5$~meV).  However, the main contribution to the optical alignment as well as to the rotation of the plane of the linear polarization and optical orientation originates from the bright exciton. We determined all main parameters of the exciton fine structure and dynamic processes in the studied NPLs. We conclude that the observation of the effects becomes possible due to the large contribution of the bright exciton to the PL intensity even at low temperatures. The presence of the CdS shell in studied CdSe/CdS NPLs results in a decrease of the electron-hole exchange interaction and corresponding decrease of the bright to dark exciton energy splitting and relaxation rate (in comparison with the CdSe NPLs without shell). We have found that the anisotropic splitting of the dark exciton states in zero magnetic field can be larger than that of the bright exciton states, that could be due to the cubic terms in the exchange interaction. It is the small anisotropic splitting of the bright exciton in studied NPLs that allowed us to observe the rotation of the plane of the linear polarization in the magnetic field in the Faraday geometry, as well as the optical orientation effect in zero magnetic field.

\section*{Methods}

\textbf{Sample fabrication.}
A set of CdSe/CdS core/shell NPLs with different shell thicknesses is studied. The fabrication procedure is described in Refs. \onlinecite{Ithurria2011,Mahler2012}. The parameters of all studied samples can be found in Table 1 of Ref. \onlinecite{Feng2020}. NPLs are passivated with oleic acid and stored in a mixed solvent consisting of 40$\%$ heptane and 60$\%$ decane. For study in cryostat, NPLs in solvent are drop-casted on Si substrate. All NPL samples were grown from the same CdSe core with an average lateral dimension of $(13.7 \pm 0.2) \times (10.8 \pm 0.2)$~nm$^2$ and a thickness of 1.2~nm (i.e., 4 monolayers). The CdSe/CdS NPLs have a total thickness of $3.8 \pm 0.5$~nm (very thin shell), $4.6 \pm 0.6$~nm (thin), $7.4 \pm 1.0$~nm (medium shell), $11.6 \pm 1.6$~nm (thick shell), $19.1 \pm 1.6$~nm (very thick shell),  including the thicknesses of the CdSe core and CdS shells on both sides. Although the effects in all samples are similar, results reported here focus on the data for the NPLs with medium shell thickness (sample number MP170214A) of 3.1~nm at each side of the core.

\textbf{Continuous wave experiment.} For polarized PL spectroscopy and Raman scattering (RS) spectroscopy the sample is placed on the variable temperature insert (1.5 - 300~K) of the helium bath cryostat with superconducting solenoid (up to 5~T). Magnetic field is applied either parallel to the light wave vector (Faraday
geometry) or perpendicular to it (Voigt geometry). For the excitation of PL and RS, a DCM dye laser is used. The laser power densities focused on the sample does not exceed 5~W/cm$^2$. PL and RS are measured in a back-scattering geometry and are analyzed by a Jobin-Yvone U-1000 double monochromator equipped with a cooled
GaAs photo-multiplier and conventional photon counting electronics. Linear and circular polarization of the PL are measured using photo-elastic modulator (PEM) in the detection path. PEM modulates circular polarization of light between $\sigma^+$ and $\sigma^-$ at a frequency of 42 kHz synchronized with the detector. Together with linear polarizer and $\lambda/4$ plate, PEM allows one to measure PL polarization $P_{l}$, $P_{l'}$, and $P_{c}$ as described in the main text.

\textbf{Time-resolved experiment.} The sample is placed into the bath cryostat with a variable temperature insert (1.5~K - 300~K) and a superconducting solenoid (up to 6~T). As the photoexcitation sources we use two semiconductor pulsed lasers with the photon energies of $2.38$~eV and $1.958$~eV, pulse duration 50 ps, and the repetition rate ranging from $0.5$~MHz to 5~MHz. The average power of the photoexcitation was controlled at 1~mW/cm$^2$. The PL was spectrally resolved by the double spectrometer with the 900~gr/mm gratings in the dispersion subtraction regime. Part of the PL band with the width of less than the 0.5~nm was detected using a photomultiplier tube designed for photon counting, and measured with time resolution with the conventional time-correlated single photon counting setup (instrumental response about 100~ps).

\textbf{ASSOCIATED CONTENT}

\textbf{Supporting Information.}

\textbf{AUTHOR INFORMATION}

Corresponding Authors: \\
Olga O. Smirnova,  Email: smirnova.olga@mail.ioffe.ru\\ 
Anna V. Rodina,  Email: anna.rodina@mail.ioffe.ru\\ 
Dmitri R. Yakovlev, Email: dmitri.yakovlev@tu-dortmund.de\\


\author{O.~O.~Smirnova}
\affiliation{\affiIOFFE}
\author{I.~V.~Kalitukha}
\affiliation{\affiIOFFE}
\author{A.~V.~Rodina}
\affiliation{\affiIOFFE}
\author{ G.~S.~Dimitriev}
\affiliation{\affiIOFFE}
\author{V.~F.~Sapega}
\affiliation{\affiIOFFE}
\author{O.~S.~Ken}
\affiliation{\affiIOFFE}
\author{V.~L.~Korenev}
\affiliation{\affiIOFFE}
\author{N.~V.~Kozyrev}
\affiliation{\affiIOFFE}
\author{S.~V.~Nekrasov}
\affiliation{\affiIOFFE}
\author{Yu.~G.~Kusrayev}
\affiliation{\affiIOFFE} 
\author{D.~R.~Yakovlev}
\affiliation{\affiE2a}
\affiliation{\affiIOFFE}
\author{B.~Dubertret}
\affiliation{	\affiParis}
\author{M.~Bayer}
\affiliation{\affiE2a}




\section*{Acknowledgements}
We thank M.M. Glazov, D.S. Smirnov, and E.L. Ivchenko for valuable discussions. We acknowledge the financial support from the Russian Foundation for Basic Research (Project 19-52-12064 NNIO-a) and the Deutsche Forschungsgemeinschaft through the International Collaborative Research Centre TRR160 (Projects B1,B2). The time-resolved measurements and their analysis by N.V.K. and O.O.S. were supported by the Russian Science Foundation (Project 20-42-01008).


\clearpage

%


%

\clearpage

\setcounter{equation}{0}
\setcounter{figure}{0}
\setcounter{table}{0}
\setcounter{page}{1}
\renewcommand{\theequation}{S\arabic{equation}}
\renewcommand{\thefigure}{S\arabic{figure}}
\renewcommand{\thepage}{S\arabic{page}}
\renewcommand{\thetable}{S\arabic{table}}

\begin{widetext}
\begin{center}

\section*{Supporting Information}

\textbf{\large Optical alignment and orientation of excitons in ensemble of core/shell CdSe/CdS colloidal nanoplatelets}\\

\vspace{\baselineskip}

{O.~O.~Smirnova$^1$, I.~V.~Kalitukha$^1$, A.~V.~Rodina$^1$, G.~S.~Dimitriev$^1$, V.~F.~Sapega$^1$, O.~S.~Ken$^1$, V.~L.~Korenev$^1$, N.~V.~Kozyrev$^1$, S.~V.~Nekrasov$^1$, Yu.~G.~Kusrayev$^1$, D.~R.~Yakovlev$^{1,2}$, B.~Dubertret$^3$, and M.~Bayer$^{2}$}
\vspace{\baselineskip}

$^1$\textit{Ioffe Institute, Russian Academy of Sciences, 194021 St.~Petersburg, Russia}

$^2$\textit{Experimentelle Physik 2, Technische Universit\"at Dortmund, 44221 Dortmund, Germany}

$^3$\textit{Laboratoire de Physique et d'\'{e}tude des
	Mat\'{e}riaux, ESPCI, CNRS, 75231 Paris, France.}

\end{center}

\subsection*{S1.Kinetics of the PL: temperature and magnetic field dependences}\label{Kin}

For square nanoplatelets in the absence of an external magnetic field, the system could be considered within the framework of a three-level model consisting of the unexcited state $|G\rangle$ and the states of the bright $|A\rangle$ and dark $|F\rangle$ exciton (Figure S1). Because of the large exchange splitting between the latter $\Delta E_{AF}$ compared to the inverse characteristic times, the polarization can be described in terms of the eigenstates populations $N_A, N_F$. The presence of splittings between the sublevels of bright $\hbar \Omega_X$ and dark $\hbar \Omega_{FX}$ excitons requires that all four levels be taken into account explicitly. During the analysis of the experimental data, it was found that the consideration of the four-level system in terms of the populations enables to describe the effect of optical alignment $P_l^l$ but does not allow one to obtain a nonzero orientation $P_c^c$ in the zero magnetic field, as well as the rotation of the linear polarization plane $P_{l'}^l$ in the magnetic field in the Faraday geometry. Effective frequency associated with the splitting between sublevels of the bright exciton turns out to be comparable with its inverse spin lifetime ($ \Omega_X T_A \sim 1$).

\begin{figure*}[h!] 
\includegraphics[width=0.95\linewidth]{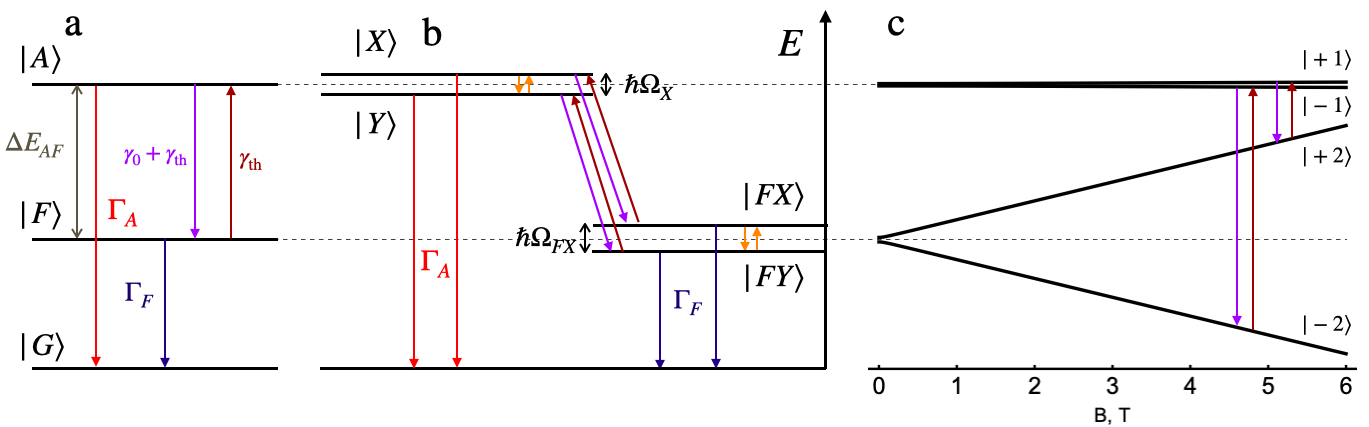}	
\caption{\label{fig:scheme} a) A scheme of the lower energy levels of the bright $ |A\rangle (\pm 1)$ and dark $|F\rangle (\pm 2)$ excitons splitted at $\Delta E_{AF}$, b) the considered transitions between states of the bright and dark excitons with characteristic rates in the absence of an external magnetic field; c) splitting of states in the magnetic field in the Faraday geometry for the maximal $g$-factors and splittings in $B=0$.}
\end{figure*}

In the main text, the system is considered in terms of pseudospins $\bm s_{A,F}$ and average pseudospins of bright and dark excitons $\bm S_{A,F}$. To describe the polarization effects under study $P_l^l, P_{l'}^l, P_{c}^c$ in terms of the total pseudospin, defined as $\bm s_{A,F} = N_{A,F}\bm S_{A,F}$, a complementary analysis of population redistribution is required. The system of equations describing the dynamic transfer of excitons between the bright and dark exciton levels, taking into account relaxation between them and recombination, can be written in the following matrix form:

\begin{equation}\label{matrixA} 
\frac{\partial}{\partial t} \begin{pmatrix}
	N_{A} \\
	N_F
\end{pmatrix} = A \begin{pmatrix}
	N_{A} \\
	N_F
\end{pmatrix} + \begin{pmatrix}
	G_{A} \\
	G_F
\end{pmatrix} , \ \ \ A = \begin{pmatrix}
	-(\Gamma_A+\gamma_0 +\gamma_{\rm th})& \gamma_{\rm th}\\
	\gamma_0+\gamma_{\rm th}&- (\Gamma_F+\gamma_{\rm th})\\
\end{pmatrix}  = \begin{pmatrix}
	-\frac{1}{\tau_A}& \gamma_{\rm th}\\
	\gamma_0+\gamma_{\rm th}&- \frac{1}{\tau_F}\\
\end{pmatrix},
\end{equation}
where $\Gamma_{A,F}$ are the radiation recombination rate of bright and dark exciton, $\gamma_0$ is the relaxation rate from the bright state to the dark one at zero temperature, $\gamma_{\rm th}$ is the thermally activated phonon-assisted relaxation $\gamma_{\rm th} = \gamma_0 N_B$, where $N_B (\Delta E_{AF})$ is the Bose–Einstein phonon occupation $N_B (E)  = 1/(\exp(E/k_B T)-1)$, $k_B$ is the Boltzmann constant. Lifetimes of the bright and dark exciton:
\begin{equation}\label{tau}
\tau_A^{-1} = \Gamma_A+\gamma_0+\gamma_{\rm th}, \ \ \ \ \tau_F^{-1} = \Gamma_F +\gamma_{\rm th}.
\end{equation}
Exciton generation is determined by $G_{A,F}$ ($G_A + G_F = 1$). We consider two ways of optical excitation: constant-wave pumping and impulse excitation. In the first case, we are interested in stationary solutions with $G_A(t), G_F(t) = \rm const$:

\begin{gather}\label{populations_cw}
N_A = \frac{\tau_A G_A + (1-\Gamma_F \tau_F) G_F} {\Gamma_F \tau_F + \Gamma_A\tau_A (1-\Gamma_F \tau_F)} \\
N_F = \frac{\tau_F (1-\Gamma_A \tau_A) G_A + \Gamma_F \tau_F}{\Gamma_F \tau_F + \Gamma_A\tau_A (1-\Gamma_F \tau_F)}.
\end{gather} 
In the quasi-resonant excitation we assume that only bright exciton with much effective oscillator strength is excited ($G_A = 1, G_F = 0$). At low temperature $\Gamma_F\tau_F = 1$. Therefore, $N_A = \tau_A G_A$, $N_F = \tau_F(1-\Gamma_A \tau_A)$. 

In the second mode the states are excited by an impulse at the initial moment of time and . For the resonant excitation we assume that only bright exciton with much effective oscillator strength is excited ($G_A = 1, G_F = 0$). In the case of non-resonant  excitation after supposedly fast relaxation from the excited level higher in energy we consider $G_A = G_F = 0.5$. The dynamics of the total photoluminescence intensity from the system in question is presented as follows:

\begin{equation}\label{intensity}
I(t) = \Gamma_A N_A (t) + \Gamma_F N_F(t) = B_1e^{-\Gamma_St}+B_2e^{-\Gamma_Lt} \ ,
\end{equation}
where $\Gamma_{S,L}$ are the eigenvalues of the matrix $A$ $\eqref{matrixA}$ and the constants $ B_1, B_2$ can be found from the initial conditions. In the absence of relaxation from the lower levels, when $\gamma_{\rm th} = 0$, that is satisfied at zero temperature, the eigenvalues are equal to the inverse of the lifetimes $\tau_A^{-1}, \tau-F^{-1}$.

Analysis of the photoluminescence kinetics makes it possible to determine some of the parameters, namely, the recombination rates of bright and dark excitons $\Gamma_A, \Gamma_F$, the energy splitting between these levels $E_{AF}$, and the relaxation rate at zero temperature $\gamma_0$. 

The expression for the fast (short) $\Gamma_{\rm S}$ and asymptotic (long) $\Gamma_{\rm L}$ decay rates of photoluminescence in the terms of the three-level model can be written as follows:

\begin{equation}\label{Gamma_As}
\Gamma_{S,L} = \frac{1}{2}\left(\Gamma_A+\Gamma_F +\gamma_0 \coth\left(\frac{\Delta E_{AF}}{2k_BT}\right) \pm \sqrt{\left( \gamma_0 +\Gamma_A -\Gamma_F \right)^2 + \gamma_0^2 \sinh^{-2}\left(\frac{\Delta E_{AF}}{2k_BT}\right)} \right) \ ,
\end{equation}

At low temperature, $\Gamma_{L} = \Gamma_F$, and at saturation, $\Gamma_{L} = \frac{\Gamma_A+\Gamma_F}{2}$. The parameters $\gamma_0, \Delta E_{AF}$ determine the bend in the temperature dependence and have some consistency. Fast component at low temperature allows one to fix the sum of rates $\Gamma_{A}+\gamma_0 = 1.11 \rm \ ns^{-1}$.

The magnetic field in the Voigt geometry mixes the bright and dark exciton states, causing additional activation of radiative recombination of the lower energy states:

\begin{equation}\label{key}
\Gamma_F = \Gamma_F(B = 0) + \left(\frac{g_e \mu_B B}{\Delta E_{AF}} \right)^2.
\end{equation}
The experimentally measured dependences of the asymptotic rate on temperature and on the magnitude of the magnetic field in the Voigt geometry are shown in Figure \ref{TRPL} (c) and (d), respectively. 

\begin{figure}[h!]
\includegraphics[width=0.8\linewidth]{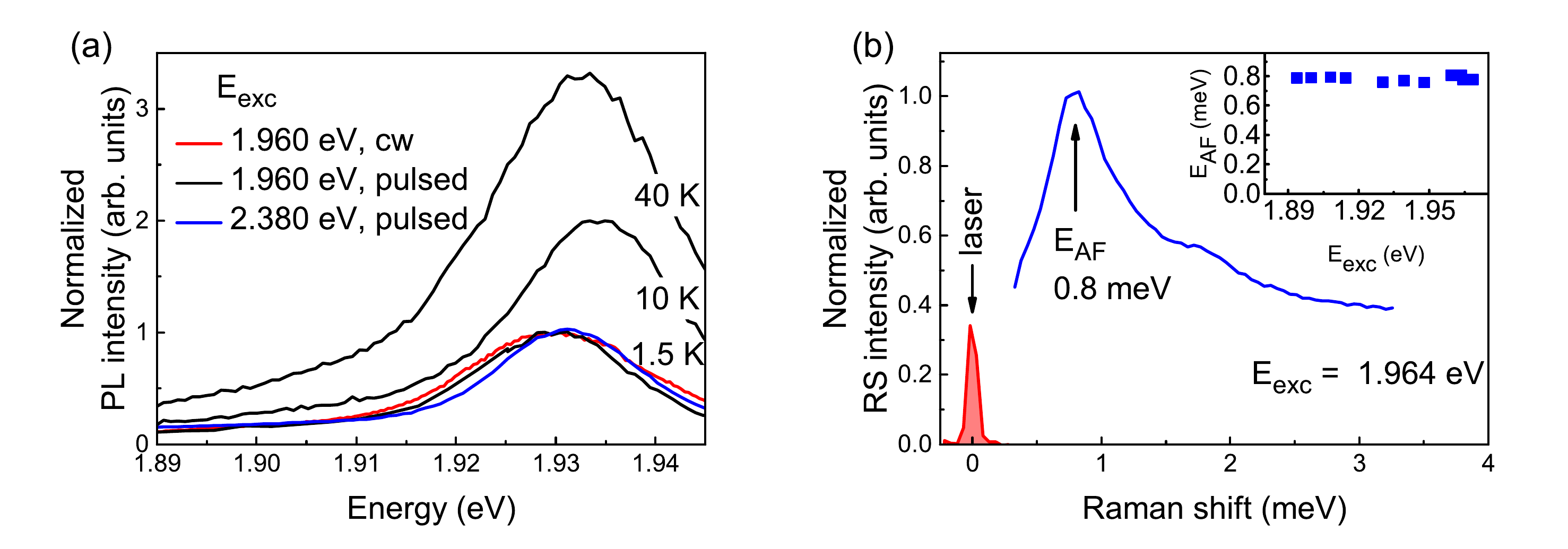}
\caption{\label{fig:spectra} PL and RS spectroscopy of the colloidal CdSe/CdS NPLs. (a) Normalized PL spectra at $T = 1.5$~K under resonant cw (red) and pulsed (black) excitation and non-resonant pulsed (blue) excitation. PL spectra (black, pulsed resonant excitation) normalized in way to conserve the ratio between PL intensities at different temperatures. (b) Raman scattering spectrum under excitation energy $E_{\rm exc} = 1.964$~eV at zero magnetic field at $T = 1.5$~K. Laser is shown by red line. Inset: The dependence of bright-dark exciton energy splitting $E_{\rm AF}$ on the excitation energy.}
\end{figure}

\begin{figure}[h!]
\includegraphics[width=0.8\linewidth]{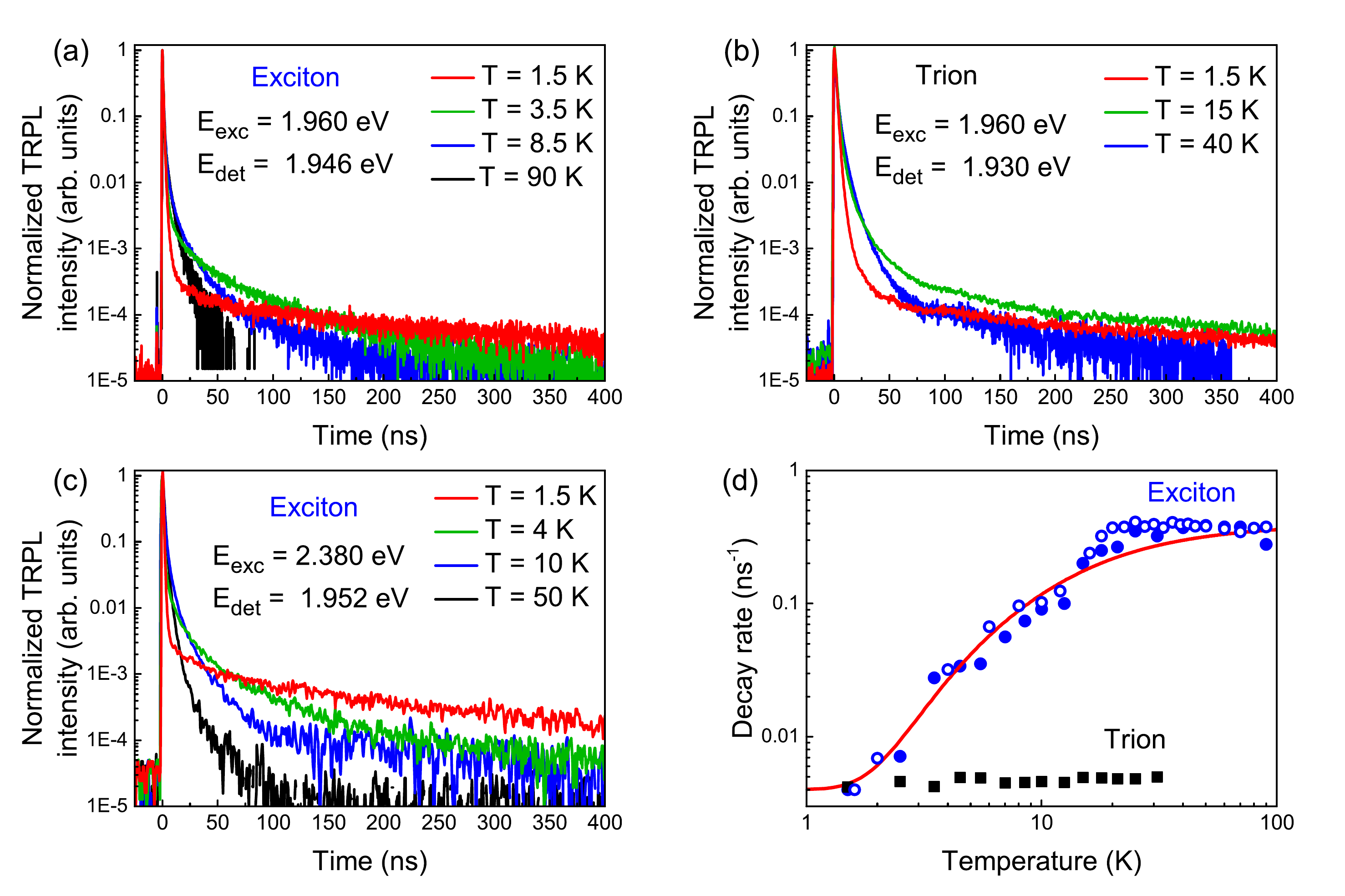}
\caption{\label{fig:TRPL_XT} Time-resolved PL spectroscopy of the colloidal CdSe/CdS NPLs in wide temperature range. (a) PL dynamics under resonant excitation detected on exciton. (b) PL dynamics under resonant excitation detected on trion. (c) PL dynamics under non-resonant excitation detected on exciton. (d) Experimental dependence of decay rate $\Gamma_{\rm L}$ on temperature for resonant (blue circles) and non-resonant (blue open circles) excitation of exciton and resonant excitation of trion (black squares). Theoretical curve is given by red line.
}
\end{figure}
The data can be coherently described by theoretical dependences for parameter values $\Gamma_A = 0.8 \rm \ ns^{-1}$, $\Gamma_F = 0.004 \rm \ ns^{-1}$, $\gamma_0 = 0.3 \rm \ ns^{-1}$, $\Delta E_{AF} = 0.8 \rm \ meV $.

\subsection*{S2. Pseudospin components}

Taking into account the relaxation time anisotropy, the expressions for the pseudospin components under arbitrary initial conditions are written as follows:

\begin{equation*}\label{Ani}
S_{AX} = \frac{S_{AX}^0(T_2 \Omega_Z^2+T_1 \Omega_X^2(1+T_2^2\Omega_A^2)- S_{AY}^0 T_2^2 \Omega_Z\Omega_A^2+S_Z^0 \Omega_X \Omega_Z(T_1(1+T_2^2\Omega_A^2)-T_2)}{\tau_A \Omega_A^2(1+T_2^2\Omega_A^2)} 
\end{equation*}

\begin{equation}\label{key}
S_{AY} = \frac{T_2(S_{AY}^0+S_{AX}^0 T_2 \Omega_Z-S_Z^0 T_2 \Omega_X)}{\tau_A(1+ T_2^2\Omega_A^2)},
\end{equation}

\begin{equation*}\label{key}
S_{AZ} = \frac{S_{Z}^0(T_2 \Omega_X^2+T_1\Omega_Z^2 (1+T_2^2\Omega_A^2)+S_{AY}^0 T_2^2 \Omega_X\Omega_A^2+S_{AX}^0 \Omega_X \Omega_Z(T_1(1+T_2^2\Omega_A^2)-T_2)  }{\tau_A \Omega_A^2(1+T_2^2\Omega_A^2)}
\end{equation*}

\subsection*{S3. Dark exciton: two extreme cases of relaxation mechanism} \label{dark}

The case of the only transferred $S_Z$ consist in two terms related to conversion of linear polarization into circular in the bright exciton and vice versa in the dark one for $P_{l}^l$ and $P_{l'}^l$. 

\begin{equation}\label{key}
	P_{lF}^l =- \frac{T_{F}^2 \Omega_{FX} \Omega_X ( \Omega_A^2 T_2^2 -\Omega_{FZ}T_F \Omega_Z(T_1(1+T_2^2\Omega_A^2)-T_2 ))}{2\tau_A \tau_F (1+\Omega_F^2 T_{F}^2)\Omega_A^2(1+\Omega_A^2 T_A^2)}
\end{equation}

\begin{equation}\label{key}
	P_{l'F}^l = \frac{T_{F}^2\Omega_{FX}\Omega_X (\Omega_A^2T_2^2 \Omega_{FZ}T_F-\Omega_Z(T_1(1+T_2^2\Omega_A^2)-T_2))}{2\tau_A \tau_F \Omega_A^2(1+\Omega_A^2 T_A^2)(1+\Omega_{F}^2 T_{F}^2)}
\end{equation}

The effect of optical orientation due to the transfer of the $S_Z$ component has one contribution:

\begin{equation}\label{key}
	P_{cF}^c = -\frac{T_{F} P_{c}^0(1+\Omega_{FZ}^2T_F^2)}{\tau_F(1+\Omega_F^2T_F^2)} \frac{(T_2 \Omega_X^2+T_1\Omega_Z^2 (1+T_2^2\Omega_A^2))}{\tau_A \Omega_A^2(1+T_2^2\Omega_A^2)}
\end{equation}

With a complete transfer of pseudospin components, there are 4 separate contributions to the experimental alignment effect: two are connected with the direct transfer of linear components, the other two are connected with the conversion of linear components in the bright exciton to circular ones, and the inverse conversion in the dark exciton. The last two terms have the same form. Contributions associated with the transfer of linear components of the pseudospin:

\begin{equation}\label{key}
	P_{lF}^l = \frac{T_{F}  P_{l}^0( \Omega_A^2 T_2+ (T_2\Omega_Z^2 +T_1 \Omega_X^2(1+\Omega_A^2))( 1 + \Omega_{FX}^2 T_{F}^2))}{2\tau_A \tau_F\Omega_A^2(1+\Omega_A^2 T_2^2)(1+\Omega_F^2 T_{F}^2)} 
\end{equation}

\begin{equation}\label{key}
	P_{lF}^l = -\frac{T_{F} T_2 P_{l}^0  \Omega_Z T_2 \Omega_{FZ} T_{F} }{\tau_A \tau_F(1+\Omega_A^2T_2^2)(1+\Omega_F^2T_{F}^2)}
\end{equation}

There are four different contributions to the rotation associated with the transfer of the linear components of the pseudospin.

\begin{equation}\label{key}
	P_{l'F}^l = \frac{T_{F} T_2 P_{l}^0 (\Omega_Z T_2 + \Omega_{FZ} T_F) }{2\tau_A \tau_F(1+\Omega_A^2 T_2^2)(1+\Omega_F^2 T_{F}^2)}
\end{equation}

\begin{equation}\label{key}
	P_{l'F}^l =  \frac{T_{F} P_{l}^0(\Omega_A^2 T_2^2\Omega_Z (1+\Omega_{FX}^2 T_{F}^2) +(T_2\Omega_Z^2 +T_1 \Omega_X^2(1+\Omega_A^2)) \Omega_{FZ}T_F)}{2\tau_A \tau_F\Omega_A^2(1+\Omega_A^2 T_2^2)(1+\Omega_F^2 T_{F}^2)}
\end{equation}

The transfer of linear components enables two more contributions due to conversions of circular polarization to linear polarization and vice versa:

\begin{equation}\label{key}
	P_{cF}^c = \frac{T_F\Omega_{FX}T_F \Omega_X(\Omega_{FZ} T_F \Omega_Z(T_1(1+T_2^2\Omega_A^2)-T_2) -\Omega_A^2T_2^2)}{\tau_F(1+\Omega_F^2T_F^2)\Omega_A^2(1+\Omega_A^2T_2^2)}
\end{equation}

\begin{figure}[h!]
	\includegraphics[width=0.7\linewidth]{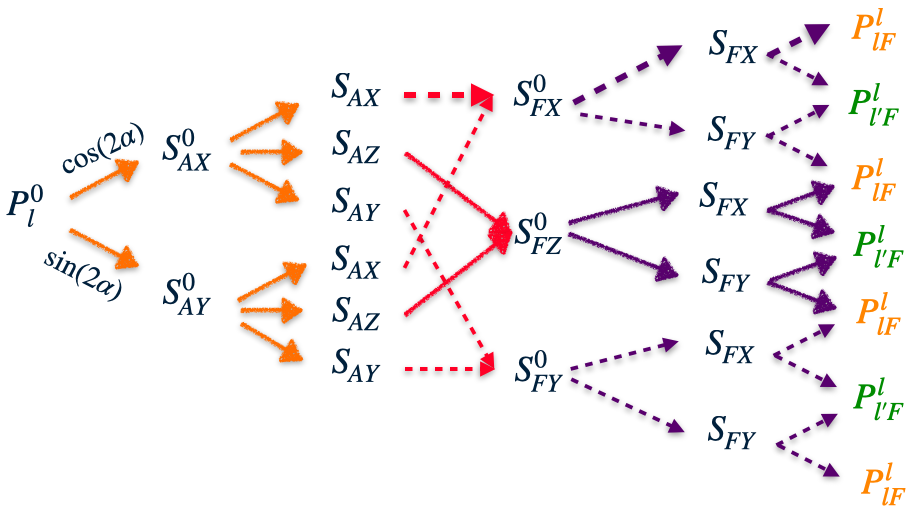}
	\caption{\label{fig:darklin} Polarization transfer scheme: transmission of linear (dashed red arrows) and circular (solid red arrows) components. In the last column the nonzero contributions to the optical alignment are marked in orange; the rotation of the linear polarization plane is marked in green.}
\end{figure}

\subsection*{S4: Analysis of the Stokes parameters in magnetic field in the Faraday geometry}

\begin{figure}[h!]
	\centering
	\includegraphics[width=0.95\linewidth]{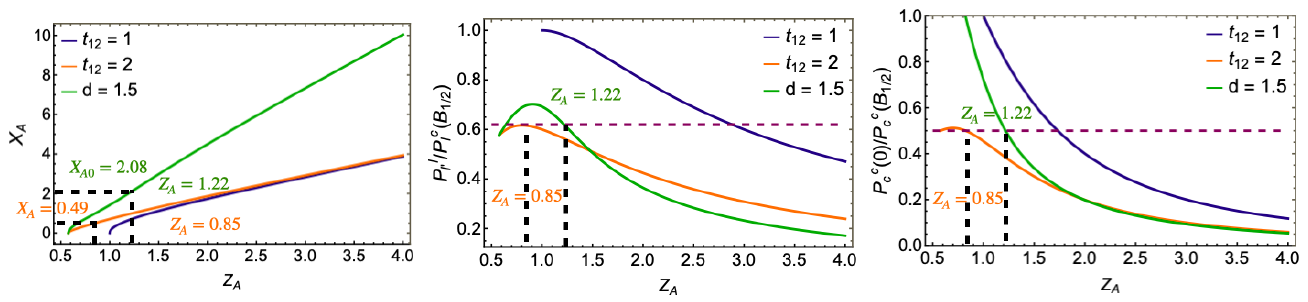}
	\caption{Three conditions on the field dependences of Stokes parameters: at $T_1 = T_2$ there is no joint solution (blue curves), at $T_1 = 2 T_2$ (orange curves) and in case of alternative analysis (see Appendix) at $ Z_A = 1.2, X_{A0}=2, d = 1.5$ a solution appears (green curves).}
	\label{fig:conditions2}
\end{figure}

\subsubsection*{Relaxation time anisotropy}
Magnetic-field conditions:

\begin{gather}\label{cond_anis}
	\frac{P_{lA}^l(B_{1/2})}{P_{lA}^l(B = 0\rm T)}= \frac{(1+X_A^2)(2Z_A^2 + X_A^2(1+t_{12}(1+X_A^2+Z_A^2)))}{(1+t_{12}(1+ X_A^2))(X_A^2+Z_A^2)(1+X_A^2+Z_A^2)} \\
	\frac{P_{l'A}^l}{P_{lA}^l} = \frac{2Z_A(X_A^2+Z_A^2)}{ 2Z_A^2 + X_A^2(1+t_{12}(1+X_A^2+Z_A^2))}, \ \ \ \ \\ \frac{P_{cA}^c(B =0 \rm T)}{P_{cA}^c(B_{1/2})} = \frac{(X_A^2+Z_A^2)(1+X_A^2+Z_A^2)}{(1+X_A^2)(X_A^2+t_{12} Z_A^2(1+X_A^2+Z_A^2))}
\end{gather}

At low temperature $I_A/(I_A+I_F) = \Gamma_A \tau_A$.  Hence, for our experiment at $T_1 =2 T_2$ we obtain the following fourth condition and parameter binding:

\begin{equation}\label{key}
{ \cal A}	P_{lA}^l = \frac{\Gamma_A P_{l}^0 T_1}{4} \left(2+ \frac{1}{1+X_A^2/4} \right) = 0.1
\end{equation}

This condition, in the case of a fixed $X_A = 1$, imposes the requirement of $P_{l}^0 T_1 = 0.18$. Further, we can define some restrictions for the parameter values. It is clear that the lifetime of the pseudospin should be less than the lifetime of the exciton $T_1, T_2 \leq \tau_A$.  At low temperature, the fast component of photoluminescence decay is determined by the lifetime of the bright exciton. The experimentally measured kinetics at $T = 1.5 K$ gives $\tau_A = 0.9 \rm \ ns$ (see Figure \ref{cw} (c)). Thus we obtain an upper limit $T_1 \leq 0.9$.  Another restriction is associated with the fact that $P_l^0\leq 1$. To get the proper amplitude with $\Gamma_{A} = 0.8$ we need to have $T_1 \geq 0.2$.  Hence, by varying $T_1$ in the range $ 0.2\rm \ ns  \leq T_1\leq 0.9 \rm \ ns$ we get valid intervals for the parameters. Possible values of the parameters are presented in the Table \ref{tab:parameters2}. 

For the amplitude of optical orientation
\begin{equation}\label{key}
{\cal A}	P_{cA}^c =   P_c^0\frac{ \Gamma_A T_2}{1+\Omega_{XA}^2T_2^2} = 0.034.
\end{equation}

From this expression we get $ P_{l}^0/ P_{c}^0 = 0.58$.  

\begin{table}[h!]
	\caption{\label{tab:parameters2}%
		Sets of bright exciton parameters in the allowed ranges at fixed ratios $T_1 =2 T_2$, $X_A = \Omega_X T_1 = 1$, $Z_A =\Omega_Z T_1= 1.7$, $\Gamma_A P_{l}^0 T_1 = 0.14$, and $ P_{c}^0/ P_{l}^0 = 0.65$, allowing description of the broad part of of the Stokes parameter dependences on magnetic field in the Faraday geometry.}
	\begin{ruledtabular}
		\begin{tabular}{dddddd}
			\multicolumn{1}{c}{\textrm{$T_1, \ \rm ns$}}&
			\multicolumn{1}{c}{\textrm{$\tau_{s1}, \ \rm ns$}}&
			\multicolumn{1}{c}{\textrm{$\tau_{s2}, \ \rm ns$}}&
			\multicolumn{1}{c}{\textrm{$P_{l}^0$}}&
			\multicolumn{1}{c}{\textrm{$g_A$}}&
			\multicolumn{1}{c}{\textrm{$\hbar \Omega_X, \ \rm \mu eV$}}\\
			\hline
			0.2& 0.26 &0.13&0.90 &0.032& 3.2\\
			0.3& 0.45 & 0.23&0.60&0.021&2.1\\
			0.4& 0.72 &0.36&0.45& 0.016& 1.6 \\
			0.5& 1.13&0.57&  0.36&0.013& 1.3 \\
			0.6&  1.8  & 0.91&0.30& 0.011& 1.1\\
			0.7& 3.15&1.59&   0.25&0.009&0.9 \\
			0.8& 7.20&3.63&   0.22&0.008&0.8 \\
			0.89& 80&40.40&   0.20&0.007&0.7 \\
		\end{tabular}
	\end{ruledtabular}
\end{table}
\begin{table}[h!]
	\caption{\label{tab:parameters3}%
		Sets of bright exciton parameters in the allowed ranges in the case of $T_A(B)$ at fixed ratios $X_{A0}= \Omega_X T_A(0)= 2.08$, $X_{AB_{1/2}}/X_{A0} =T_A(B_{1/2})/T_A(0)= 2/3$, $Z_A=\Omega_Z T_A(B_{1/2}) = 1.22$, allowing description of the broad part of of the Stokes parameter dependences on magnetic field in the Faraday geometry.}
	\begin{ruledtabular}
		\begin{tabular}{dddddd}
			\multicolumn{1}{c}{\textrm{$T_A(0), \ \rm ns$}}&
			\multicolumn{1}{c}{\textrm{$\tau_{sA}(0), \ \rm ns$  }}&
			\multicolumn{1}{c}{\textrm{$\tau_{sA}(B_{1/2}), \ \rm ns$ }}&
			\multicolumn{1}{c}{\textrm{ $P_{l}^0$}}&
			\multicolumn{1}{c}{\textrm{$g_A$}}&
			\multicolumn{1}{c}{\textrm{$\hbar \Omega_X, \ \rm \mu eV$}}\\
			\hline
			0.22& 0.31& 0.18&  0.96& 0.031 & 6.2  \\
			0.3&  0.45& 0.26& 0.70&0.023& 4.6 \\
			0.4&  0.72& 0.38&0.53& 0.017 & 3.4 \\
			0.5&   1.1& 0.53&0.42&0.014& 2.7 \\
			0.6&  1.8& 0.72&0.35 &0.012& 2.3 \\
			0.7&3.2& 0.97&0.30 & 0.010& 2.0 \\
			0.8&7.2& 1.3&0.26 & 0.009& 1.7\\
			0.89&80& 1.7&0.24 & 0.008& 1.5\\
		\end{tabular}
	\end{ruledtabular}
\end{table}

Then to extract the parameters of the dark exciton we can use the amplitude of the narrow contour at low temperature:
\begin{equation}\label{amplF}
{\cal F} P_{lF}^l(0) = \gamma_0 \frac{T_F T_1 P_l^0}{2\tau_F}  = 0.014.
\end{equation}

Here we used the fact that at low temperature $I_F/(I_A+I_F) = \gamma_0 \tau_A$. The long component of photoluminescence decay at $T = 1.5 K$ is determined by the lifetime of the dark exciton. The kinetics gives $\tau_F = 250 \rm \ ns$ (see S1). As for $\gamma_0$, we can estimate it from low-temperature bright exciton lifetime $\tau_A = (\Gamma_A+\gamma_0)^-1 = 0.9 \rm \ ns$ with fixed $\Gamma_A = 0.8 \rm ns^{-1}$: $\gamma_0 = 0.3 \rm ns^{-1}$. 

For our range of $g_A$ we get $g_F \in [3.31, 3.358]$. Therefore $\hbar \Omega_{FX} = (11.55 \pm 0.05 )\ \rm \mu eV$. The fact that the splitting between the dark exciton states in the absence of a magnetic field turned out to be larger than the anisotropic splitting between the bright exciton states does not contradict the problem conditions, since they have a different nature. The splitting between the states of the bright exciton is driven by the anisotropy of the nanoplatelets in the plane. For the dark exciton, the cubic by $J$ term of the Hamiltonian also leads to the splitting of its states.

\subsubsection*{Magnetic field dependence of the relaxation}

In terms of $X_{A} = \Omega_X T_A(0) $ and $Z_A = \Omega_Z T_A(B_{1/2})$ we can introduce a parameter that is responsible for the variation of the pseudospin lifetime in the external magnetic field:
\begin{equation}\label{key}
\frac{1}{d} = \frac{X_{AB_{1/2}}}{X_{A}} = \frac{T_A(B_{1/2})}{T_A(0)}
\end{equation}
The conditions on the magnetic field dependences can be rewritten as follows:
\begin{equation}\label{Conditionsiii}
\frac{P_{lA}^l(B_{1/2})}{P_{lA}^l(0)}= \frac{1}{d} \frac{2d^2+ X_{A}^2}{ X_{A}^2 +d^2(1+ Z_A^2)} \frac{1+X_{A}^2}{2+X_{A}^2}= \frac{1}{2},\ \ \ \  \frac{P_{l'A}^l(B_{1/2})}{P_{lA}^l(B_{1/2})} = \frac{2d^2Z_A}{2d^2+ X_{A}^2}, \ \ \ \frac{P_{cA}^c(0)}{P_{cA}^c(B_{1/2})} = \frac{1}{d}\frac{X_{A}^2+d^2(1+Z_A^2)}{(1+X_{A}^2)(1+Z_A^2)}.
\end{equation}

For the given experimental values, the solution is the set of parameters: $ Z_A = 1.2, X_{A}=2, d = 1.5$, what can be seen graphically in Figure S5 in green.

The amplitude of optical alignment has the form:

\begin{equation}\label{key}
P_{lA}^l = \frac{\Gamma_A P_l^0 T_A(0)}{2}\frac{2+X_{A0}^2}{1+X_{A0}^2} = 0.1.
\end{equation}
From this we get $\Gamma_A S_{AX}^0 T_A(0) = 0.084$. With the analogous restrictions $\Gamma_A \leq 1/\tau_A = 1.33 \ \rm ns^{-1}$, $S_{AX}^0\leq 0.5$ for the pseudospin $T_A$ we obtain $0.13\leq T_A(0) \leq 0.75$. Sets of the obtained bright exciton parameters are presented in the Table \ref{tab:parameters3}.

\end{widetext}

\end{document}